\documentclass[aps,twocolumn,showpacs,preprintnumbers,amsmath,amssymb,pre]{revtex4-1}
\usepackage{epsf,amsmath,amssymb,amsfonts,verbatim,color,multirow,pifont}
\usepackage{graphicx}
\usepackage{bm}% textbf math
\usepackage{dcolumn}% Align table columns on decimal point
\usepackage{txfonts}
\usepackage{hyperref}
\usepackage{soul}

\begin{document}
\title{Effects of the non-Markovianity and non-Gaussianity of active environmental noises on engine performance}
\author{Jae Sung Lee$^{1}$} \email{jslee@kias.re.kr}
\author{Hyunggyu~Park$^{1}$} \email{hgpark@kias.re.kr}
\affiliation{{$^1$School of Physics and Quantum Universe Center, Korea Institute for
Advanced Study, Seoul 02455, Korea}}

\newcommand{\revise}[1]{{\color{red}#1}}

\date{\today}

\begin{abstract}
An active environment is a reservoir containing \emph{active} materials, such as bacteria and Janus particles. Given the self-propelled motion of these materials, powered by chemical energy, an active environment has unique, nonequilibrium environmental noise. Recently, studies on engines that harvest energy from active environments have attracted a great deal of attention because the theoretical and experimental findings indicate that these engines outperform conventional ones. Studies have explored the features of active environments essential for outperformance, such as the non-Gaussian or non-Markovian nature of the active noise. However, these features have not yet been systematically investigated in a general setting. Therefore, we systematically study the effects of the non-Gaussianity and non-Markovianity of active noise on engine performance. We show that non-Gaussianity is irrelevant to the performance of an engine driven by {any linear force (including a harmonic trap) regardless of time dependency}, whereas non-Markovianity is relevant. However, for a system driven by a general nonlinear force, both non-Gaussianity and non-Markovianity enhance engine performance. Also, the memory effect of an active reservoir should be considered when fabricating a cyclic engine.
\end{abstract}

%\pacs{05.70.-a, 05.40.-a, 05.70.Ln, 02.50.-r}

\maketitle

\section{Introduction}
\label{sec:intro}
Thermodynamics is the study of \emph{open} systems, i.e., systems that are not isolated, instead interacting with the environment. Thus, the dynamics of an open system are significantly affected by the characteristics of the environment. For conventional thermodynamic problems, the environment has been assumed to be in thermal equilibrium. In such an environment, fluctuations of the system observable and energy dissipation are regulated by the fluctuation-dissipation theorem (FDT)~\cite{Risken}. Using the Langevin terminology, for simplicity the equilibrium noise is assumed to be Gaussian and typically Markovian. Recent important discoveries in thermodynamics, such as fluctuation theorems~\cite{Seifert2005, Jarzynski1997, Crooks1999, LeeHK2013, Sagawa2012, Noh2012, JSLee2018} and thermodynamic uncertainty relationships~\cite{Barato2015, Gingrich2016, Hasegawa, Dechant, Koyuk, JSLee2019, JSLee2021}, were made with the environment assumed to be in equilibrium.

Over the last two decades, thermodynamics has been revisited using \emph{active} environments, i.e., reservoirs containing active particles such as bacteria and Janus particles~\cite{Kanazawa, Wu2000, Maggi2014, Krishnamurthy, Marconi2017}. As the active particles are self-propelled by chemical energy, an active reservoir is intrinsically in a nonequilibrium state. In an active environment, the FDT does not apply; noise differs from an equilibrium environment. Usually, the noise can be either non-Gaussian or non-Markovian with the FDT violation~\cite{Kanazawa, Krishnamurthy, Leptos, Kurtuldu}. This has provoked vigorous debate on how to establish the thermodynamics of a system in contact with an active environment; it is necessary to appropriately define key thermodynamic quantities including heat, entropy, and temperature~\cite{Marconi2017, Fodor, Mandal, Dabelow}.

The nonequilibrium characteristics of an active environment make it possible to design novel microscopic motors or engines. As an example, the autonomous engine operates in the absence of an external driving force. When an asymmetric \emph{passive}, (i.e., not self-propelled) object interacts with an active reservoir, a directional current is generated by the asymmetric passive object or active particles ~\cite{Sokolov2010, Leonardo, Angelani, Vizsnyiczai, Galajda}. This is unprecedented; the unidirectional motion can be exploited to design a microscopic motor that works in an active environment. Recently, it was shown that the reverse was also possible; an asymmetric active particle immersed in an equilibrium reservoir can produce a unidirectional current~\cite{Aubret, Kummel, JSLee2021-2}.

The next example, which is our main focus in this study, is an engine operated by an external driving force. In Ref.~\cite{Krishnamurthy}, a microscopic Stirling engine was realized in a bacterial reservoir experimentally. The performance was better than that of a conventional engine working in an equilibrium reservoir. This discovery prompted several studies on the cause of the outperformance. It was suggested in Ref.~\cite{Krishnamurthy} that the non-Gaussian nature of the active noise might explain the performance enhancement, but it was later shown that non-Gaussianity per se did not affect the performance of the Stirling engine~\cite{entropy}. It was also shown that the efficiency of a Brownian heat engine was enhanced by non-Markovian but Gaussian active noise~\cite{JSLee2020active}.

These studies raise an obvious question: which features of active noise are relevant to engine performance enhancement in a general setting? We systematically study the effect of the non-Markovianity and non-Gaussianity of general active environmental noise on engine performance within active reservoirs. It is easy to show that non-Gaussianity is irrelevant to engine performance in a system  { driven by any \emph{linear} force (including a harmonic trap) regardless of time dependency}. This is consistent with the result of Ref.~\cite{entropy}; thus, non-Markovianity explains the outperformance noted in Ref.~\cite{Krishnamurthy}. If there is a non-harmonic potential (or nonlinear external force), non-Gaussianity also affects engine performance. We calculated the work and heat of an engine driven by general external forces with various types of active noise. We also studied cyclic engines within active reservoirs that change periodically, and found that the memory effect of an active reservoir should be considered when investigating engine performance.

This paper is organized as follows. In Sec.~\ref{sec:various_models}, we review the various active-noise models; these include shot noise, colored-Poisson noise, the active Ornstein-Uhlenbeck process and active Brownian particle models. In Sec.~\ref{sec:first_law}, we discuss the first law of thermodynamics for active systems. In Sec.~\ref{sec:linear}, the effects of non-Markovianity and non-Gaussianity on engine performance are discussed when the engine is driven by a linear force (a harmonic potential). In Sec.~\ref{sec:nonlinear}, we study the effect of a nonlinear force. We conclude the paper in Sec.~\ref{sec:conclusion}.

\section{Various Models for Active Noise}
\label{sec:various_models}
We consider a $N$-dimensional Brownian particle (or one-dimensional $N$ particles) immersed in active reservoirs. The stochastic motion of the Brownian particle is induced by the interaction with the active reservoirs. This stochastic motion and the interaction can be phenomenologically  described by the following overdamped Langevin equation~\cite{Fodor, Mandal, entropy, JSLee2020active}:
\begin{equation}
	\gamma_i \dot{x}_i = f_i(\textbf{\textit{x}},t) + \gamma_i \zeta _i~~~(i=1, \cdots, N)
	\label{eq:overLangevin}
\end{equation}
where $\textbf{\textit{x}}=(x_1, \cdots, x_N)^\textsf{T}$ is the position of the particle, $\gamma_i$ is a dissipation coefficient, $f_i (\textbf{\textit{x}}, t)$ is an external force at time $t$, and $\zeta_i$ describes a random noise from an active reservoir.

Depending on the statistics of $\zeta_i$, various `active-noise' models exist~\cite{Kanazawa, Fodor, Mandal,  entropy, JSLee2020active, Pak, ABP1, ABP2}. One common feature of the models is that the autocorrelation function of the noise exhibits an exponentially decaying behavior in time difference as follows:
\begin{align}
	\langle \zeta_i(t) \zeta_j(t^\prime) \rangle = \delta_{ij} \gamma_i^{-2}  \frac{ D_i}{ \tau_i } e^{-\frac{|t-t^\prime|}{\tau_i}},  \label{eq:corr}
\end{align}
where $D_i$ is the noise strength, $\tau_i$ is the persistence time, and the average noise is given by $\langle \zeta_i (t) \rangle = 0$.
Non-Markovianity is quantified by finite $\tau_i$ and the $\delta$-correlated white noise is obtained in the $\tau_i \rightarrow 0$ limit.
Note that  the FDT is not satisfied for finite $\tau_i$ in Eq.~\eqref{eq:overLangevin} in contrast to the standard generalized Lengevin equation, thus yielding a nonequilibrium steady state.

Nonequilibrium noise is characterized not only by non-Markovianity but also by non-Gaussianity. Non-Gaussianity of a noise can be simply checked from nonzero higher order (more than second order)  cumulants of a noise. In the following subsections, we introduce four different active-noise models; shot noise, colored-Poisson noise, active Ornstein-Uhlenbeck process (AOUP), and active Brownian particle (ABP) noise models, which  exhibit distinct features in terms of non-Markovianity and non-Gaussianity. For example, the AOUP noise is Gaussian, while
the ABP is non-Gaussian. The colored-Poisson noise is neither Gaussian nor Markovian, but their nonequilibrium features can be controlled systematically by varying noise parameters. The shot noise is obtained in the zero-persistence time limit of the colored-Poisson noise.

It is worthy to mention that there may be additional noises originated from passive particles like water molecules in the surrounding medium. In this more realistic situation, one should add a Gaussian white noise to Eq.~\eqref{eq:overLangevin} as done in Refs.~\cite{Dabelow,Pak}. However, such an addition does not change our main conclusion, thus we focus here the case only with an active noise
for simplicity.

\subsection{shot (white-Poisson) noise model}
\label{sec:shot}

First, we consider the `shot'-noise or `white Poisson'-noise  model~\cite{Kanazawa,entropy}.  In this model, $\zeta_i (t)$ is given by
\begin{align}
\zeta_i (t)=  \sum_n c_{i,n} \delta(t-t_{i,n}),	\label{eq:Shot_noise}
\end{align}
where  $c_{i,n}$ is the noise magnitude determined by a given distribution $p_i(c)$ and $t_{i,n}$ is the $n$-th event time of the Possion process with rate $\lambda_i$.  Note that the time interval of successive Poisson events ($\Delta t_{i,n} = t_{i,n+1} - t_{i,n}$) obeys the distribution $P(\Delta t) = \lambda e^{-\lambda \Delta t}$. The $\delta$-function type impulse of this noise describes a sequence of microscopic discrete events such as random collisions of bacteria~\cite{Kanazawa, Pak} without any memory in the first
approximation~\cite{entropy}. Thus, this shot noise is Markovian with  $\tau_i \rightarrow 0$ in Eq.~\eqref{eq:corr} and its autocorrelation function is given by~\cite{Kanazawa,entropy}
\begin{align}
	\langle \zeta_i(t) \zeta_j (t^\prime) \rangle = \delta_{ij}
	\lambda_i \langle c^2 \rangle_{p_i} \delta (t-t^\prime), \label{eq:Shot_corr}
\end{align}
where $\langle \cdots \rangle_{p_i}$ denotes average over the distribution $p_i(c)$ with  $\langle c_{i,n} \rangle_{p_i} =0$.
%, otherwise the Brownian particle would move unidirectionally without applying an external force.
Though Eq.~\eqref{eq:Shot_corr} shows the same $\delta$-correlated property as that of the equilibrium white noise, the shot noise leads to a nonequilibrium  steady state as shown in Fig.~\ref{fig:distribution}(b)  due to the discrete nature of the noise. The equilibrium (Gaussian white) limit is obtained  by taking the $\lambda_i \rightarrow \infty$ limit with keeping the noise strength constant as $\lambda_i  \langle c^2 \rangle_{p_i} =  2 D_i /\gamma_i^2 $~\cite{Kanazawa}. In this limit, we can infer an effective temperature from the noise strength as $D_i /\gamma_i \equiv T_i$ in the Boltzmann constant unit by setting $k_\textrm{B} =1$. For finite $\lambda_i$ and $\langle c^2 \rangle_{p_i}$, the shot noise is Markovian, but non-Gaussian, which can be checked from the nonzero fourth cumulant of the noise~\cite{entropy}.

\subsection{colored-Poisson noise model}
The colored-Poisson noise is a generalized version of the white-Poisson noise~\cite{Pak}. In this model, $\zeta_i (t)$ is given by
\begin{align}
	\zeta_i (t)=  \sum_n \frac{c_{i,n}}{\tau_i} H(t-t_{i,n}) e^{-\frac{t-t_{i,n}}{\tau_i}}, \label{eq:CPoisson_noise}
\end{align}
where $c_{i,n}$ and  $t_{i,n}$ are defined in Eq.~\eqref{eq:Shot_noise}, and $H(t)$ is the Heaviside step function:
$H(t)=1$ for $t>0$, $0$ for $t<0$, and $1/2$ at $t=0$. With this noise, each $n$-th impulse from the collision at time $t_{i,n}$ exponentially decays with the finite persistence time $\tau_i$. Thus, this noise is non-Markovian and the Markovian white-Poisson noise is obtained in the $\tau_i \rightarrow 0$ limit. The noise autocorrelation function is
\begin{align}
	\langle \zeta_i(t) \zeta_j (t^\prime) \rangle = \delta_{ij}
	\frac{\lambda_i \langle c^2 \rangle_{p_i} }{2  \tau_i}  e^{-\frac{|t-t^\prime|}{\tau_i}}, \label{eq:Colored_corr}
\end{align}
which is explicitly derived in Appendix \ref{seca:autocorr_CPoi}.
By comparing Eq.~\eqref{eq:Colored_corr} with  Eq.~\eqref{eq:corr}, we can identify $D_i = \gamma_i^2 \lambda_i \langle c^2  \rangle_{p_i} /2$.

The non-Gaussianity of this noise can be checked from the calculation of the fourth cumulant of the noise. For a Poisson noise $\zeta (t)$ with the form $
\zeta (t) =  \sum_n  c_{n} H(t-t_{n}) h(t-t_{n}) $, where $h(t)$ is an arbitrary function of time, its fourth cumulant is given by~\cite{Manuel}
\begin{align}
	\langle \langle \zeta(t)^4 \rangle \rangle = \lambda \langle c^4 \rangle_p \int_0^t dt^\prime~ [h(t-t^\prime)]^4,
\end{align}
where $\langle \langle \cdots \rangle \rangle$ stands for the cumulant average. Thus, the fourth cumulant of  the colored-Poisson noise
with $h(t)=e^{-t/\tau}/\tau$ is given by $\lambda \langle c^4 \rangle_{p_i} /(4 \tau^3)$, which may serve as a measure for the non-Gaussianity of the noise. Hence, the colored-Poisson noise model provides a systematic way to study the effect of the non-Markovianity and the non-Gaussianity of an active noise by controlling the two parameters, $\lambda$ and $\tau$.

\subsection{active Ornstein-Uhlenbeck Process (AOUP) model}
In this model, the noise $\zeta_i (t)$ satisfies the following Ornstein-Uhlenbeck process:
\begin{align}
	\tau_i \dot{\zeta}_i = -\zeta_i + \sqrt{2 D_i/\gamma_i^2} \xi_i, \label{eq:AOUP}
\end{align}
where $\xi_i $ is a Gaussian white noise with zero mean and unit variance. It is straightforward to see that the steady state distribution of $\zeta_i$ is Gaussian from Eq.~\eqref{eq:AOUP}. Thus, the AOUP noise is Gaussian. Using the general solution of Eq.~\eqref{eq:AOUP}, $\zeta_i (t) = e^{-t/\tau_i} \zeta_i (0) + \sqrt{2D_i/\gamma_i^2} \tau_i^{-1} \int_0^t e^{-(t-s)/\tau_i} \xi_i (s) ds$, the noise autocorrelation can be calcuated as
\begin{align}
	\langle \zeta_i(t) \zeta_j(t^\prime) \rangle =& e^{-\frac{t}{\tau_i}-\frac{t^\prime}{\tau_j}} \zeta_i(0) \zeta_j(0) \nonumber \\
	&+ \delta_{ij} \gamma_i^{-2}\frac{D_i}{\tau_i} \left( e^{-\frac{|t-t^\prime|}{\tau_i}} - e^{-\frac{t+t^\prime}{\tau_i}} \right).  \label{eq:AOUPcorr}
\end{align}
In the steady state where $t,t^\prime \gg \tau_i, \tau_j$, the autocorrelation function takes the same form in  Eq.~\eqref{eq:corr}. Thus, the AOUP noise is non-Markovian, but Gaussian. The Gaussian white noise is obtained in the $\tau_i \rightarrow 0$ limit.

% % % % % % % % % % % % % % % % % %
\begin{figure*}
	\centering
	\includegraphics[width=0.99\linewidth]{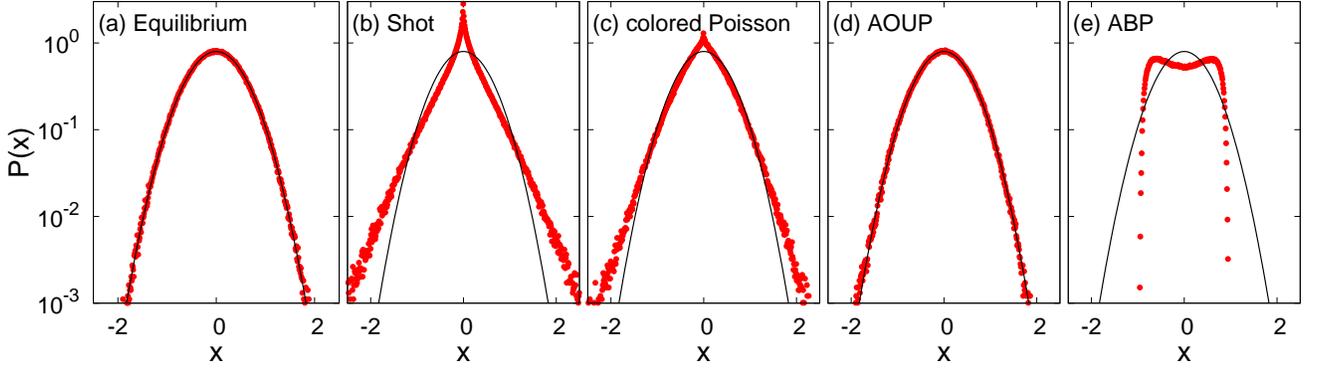}
	\caption{Steady-state distributions of a one-dimensional particle trapped in a harmonic potential for various environmental noises. For all distributions, $\langle x^2 \rangle =1/4$. The solid curve denotes the Gaussian distribution. The distributions are non-Gaussian except for the equilibrium and the AOUP noise. Parameters used for these simulations are as follows: $\lambda =1/2$ and $\langle c^2 \rangle_p =1$ are used for the shot-noise model, $\tau = 1$, $\lambda = 1$, and $\langle c^2\rangle_p=1$ are used for the colored-Poisson noise model, $D=1/2$ and $ \tau=1$  are used for the AOUP model, and $v_0 =1$ and $\tau = 1$ are used for the ABP model.
	} \label{fig:distribution}
\end{figure*}
% % % % % % % % % % % % % % % % % % % % %

\subsection{active Brownian particle (ABP) model}

Consider a self-propelled particle moving in a two-dimensional space.
The ABP model describes the motion of the active particle with the self-propulsion speed $v_0$~\cite{ABP1,ABP2}. Its motion can be described by the following overdamped Langevin equation:
\begin{equation}
	\gamma \dot{\textbf{\textit{x}}} = \gamma v_0 \textbf{\textit{e}}_\theta + \textbf{\textit{f}} (\textbf{\textit{x}},t) +\sqrt{2\gamma T}{\bm \xi},~~~\dot{\theta} = \sqrt{2 D_\theta} \xi_\theta
	\label{eq:overLangevinABP}
\end{equation}
where $ \textbf{\textit{e}}_\theta = (\cos\theta,\sin\theta)^\textsf{T}$ is the unit self-propulsion vector, ${\bm \xi}=(\xi_1,\xi_2)^\textsf{T}$ and $\xi_\theta$ are the Gaussian white noises with zero mean and unit variance, $\textbf{\textit{f}} (\textbf{\textit{x}},t)= (f_1(\textbf{\textit{x}},t), f_2(\textbf{\textit{x}},t))^\textsf{T}$ is an external force, and  $D_\theta$ is the noise strength for the rotational angle $\theta$. If we ignore the equilibrium noise $\sqrt{2\gamma T}{\bm \xi}$, regard the self-propulsion term as an active noise ${\bm \zeta} \equiv  v_0 \textbf{\textit{e}}_\theta$, and integrate the angular dynamics, we obtain the same equation as Eq.~\eqref{eq:overLangevin}. Then the noise autocorrelation function becomes~\cite{ABP1}
\begin{align}
	\langle \zeta_i (t) \zeta_j (t^\prime) \rangle_\theta = \delta_{ij} \frac{ v_0^2}{2} e^{-\frac{| t_2 - t_1 |}{\tau}}, \label{eq:ABPcorr}
\end{align}
where $\tau = 1/D_\theta$ and $\langle \cdots \rangle_\theta$ denotes an average over the $\theta$ variable. By comparing Eq.~\eqref{eq:ABPcorr} with Eq.~\eqref{eq:corr}, we can identify $D= \gamma^2 v_0^2 \tau/2$. For completeness, we present the derivation of Eq.~\eqref{eq:ABPcorr} in {Appendix \ref{seca:autocorr_ABP}}. Note that the noise autocorrelation can be also calculated in the three dimensions~\cite{ABP2}.

\subsection{steady states of various active-noise models}
Different active-noise models yield different steady states. To illustrate the difference, we perform numerical simulations of the one-dimensional Brownian motion trapped in a harmonic potential, $f(x)=-kx$ with $k=1$ and $\gamma=1$, described by Eq.~\eqref{eq:overLangevin}. From the simulations, we obtain the steady-state probability distribution of $x$ for various noise models from $10^6$ data, which are shown in Fig.~\ref{fig:distribution}. For all these simulations, we set the parameters for the second moment of $x$ being $\langle x^2 \rangle =1/4$. Parameters of respective models are specified in the caption of Fig.~\ref{fig:distribution}.

Figure~\ref{fig:distribution} (a) shows the steady-state distribution when $\zeta$ is an equilibrium noise, i.e., the Gaussian white noise satisfying $\langle \zeta(t) \zeta(t^\prime) \rangle = 2D \delta(t-t^\prime) $ with temperature $T = D/\gamma = 1/4$. As expected, the distribution is exactly Gaussian.
The steady-state distribution of the shot-noise model is evidently deviated from the Gaussian as shown in Fig.~\ref{fig:distribution} (b), even though the noise autocorrelation function and the second moment of $x$ are the same as those of the equilibrium noise.  Figure~\ref{fig:distribution} (c) shows the steady-state distribution of the colored-Poisson noise model, which is also non-Gaussian.
Different from the other active-noise models, the AOUP model results in the Gaussian distribution as shown in Fig.~\ref{fig:distribution} (d), which is due to the Gaussianity of the noise. Finally, Fig.~\ref{fig:distribution} (e) is
the steady-state distribution {of the two-dimensional ABP model along the $x$ direction}. The distribution has two symmetric peaks, which have been usually observed in the ABP systems~\cite{ABP_bimodal}.

\section{Extended thermodynamic first law} \label{sec:first_law}

To investigate thermodynamic properties of active systems, it is important to understand the thermodynamic laws governing the dynamics. In this section, we discuss  the thermodynamic first law of active systems, which  is essentially the energy conservation relation between work, heat, and system energy. Before going into active systems, we first briefly review the thermodynamic first law for a stochastic systems with equilibrium baths.

Consider an overdamped Brownian particle driven by a nonconservative force $\textit{\textbf{f}}^\textrm{nc} (\textit{\textbf{x}})$ and a conservative force $ - \nabla U(\textit{\textbf{x}}, \lambda )$, where $\lambda = \lambda(t)$ denotes a time-dependent protocol. The $i$-th degree of freedom of the particle is in contact with the equilibrium bath with temperature $T_i$. Then, its equation of motion is
\begin{align}
	\Gamma \dot{\textit{\textbf{x}}} =   - \nabla U(\textit{\textbf{x}}, \lambda )+ \textit{\textbf{f}}^\textrm{nc} (\textit{\textbf{x}}) + \Gamma \bm{\zeta}, \label{eq:Langevin_equil}
\end{align}
where the dissipation matrix $\Gamma _{ij}= \delta_{ij} \gamma_i$ and $ \bm{\zeta}$ is Gaussian white noise satisfying $\langle \zeta_i (t) \zeta_j (t^\prime) \rangle = \delta_{ij} 2T_i \gamma_i^{-1} \delta(t-t^\prime)$. Multiplying $\dot{\textit{\textbf{x}}}dt$ to Eq.~\eqref{eq:Langevin_equil} and using the chain rule $dU = \nabla U \circ \dot{\textit{\textbf{x}}} dt + \partial_\lambda U \dot{\lambda} dt $, we obtain the following thermodynamic first law~\cite{Sekimoto}:
\begin{align}
	dU = dW^\textrm{p} + dW^\textrm{nc} + \sum_i dQ_i, \label{eq:first_law}
\end{align}
where heat $dQ_i$ from bath $i$, work done by the protocol $dW^\textrm{p}$ (called as Jarzynski work~\cite{Jarzynski1997}), and work done by the nonconservative force $dW^\textrm{nc}$ are defined as
\begin{subequations}
	\begin{align}
	&dQ_i = (-\gamma_i \dot{x}_i +\gamma_i \zeta_i)\circ \dot{x}_i dt, \label{eq:heat_work_def1} \\
	&dW^\textrm{p} =  \partial_\lambda U \dot{\lambda} dt, \label{eq:heat_work_def2} \\
	&dW^\textrm{nc} = \dot{\textit{\textbf{x}}}^\textsf{T} \circ \textit{\textbf{f}}^\textrm{nc} (\textit{\textbf{x}})   dt. \label{eq:heat_work_def3}
	\end{align}
\end{subequations}
Here $\circ$ denotes the Stratonovich product.
%Note that total work $W$  is the sum of conservative and nonconservative parts as $W = W^\textrm{p} + W^\textrm{nc}$ and $ W^\textrm{c}$ is sometimes called Jarzynski work~\cite{Jarzynski1997}.

For the thermodynamic first law of active systems being set up, the work, heat, and system energy should be identified first. Defining the system energy and work are straightforward: The same definitions, Eqs.~\eqref{eq:first_law}, \eqref{eq:heat_work_def2}, and \eqref{eq:heat_work_def3} are acceptable for an active system without ambiguity.
The standard definition for `heat' is simply given by Eq.~\eqref{eq:heat_work_def1} even for the active noise~\cite{entropy, JSLee2020active, Holubec2020}, which represents the work done by the general environment (energy transfer from the active reservoir).
One may also consider the `housekeeping' heat for maintaining the nonequilibrium steady state of an active bath and thus introduce
an extra dynamics of active particles in the active bath to calculate the housekeeping heat dissipation~\cite{Fodor2021}.
Though this approach can deal with some part of dissipation of active particles, full dissipation produced from complicated chemical and mechanical operations occurred inside and around active particles still cannot be taken into consideration. Moreover, the simple thermodynamic first law in Eq.~\eqref{eq:first_law} is not applicable to this approach. Here, we take the standard definition of heat in Eq.~\eqref{eq:heat_work_def1} for simplicity and generality.

\section{Engine with a Linear force}
\label{sec:linear}

\subsection{work and heat for a linear system}
\label{sec:linearA}

First, we consider the case where the total mechanical force is given as a linear force in position such as
\begin{align}
\textbf{\textit{f}}(\textbf{\textit{x}}) = - \Gamma \textsf{A}(t)~ \textbf{\textit{x}}~,
\end{align}
where $\textsf{A}(t)$ is a time-dependent force matrix. Then, Eq.~\eqref{eq:overLangevin} can be written as
\begin{align}
	\dot{\textbf{\textit{x}}} = - \textsf{A}(t) ~\textbf{\textit{x}} + \bm{\zeta}, \label{eq:linearEq}
\end{align}
where
$\bm{\zeta} = (\zeta_1, \cdots, \zeta_N)^\textsf{T}$. Note that the force matrix can be divided into the conservative and nonconservative part as  $\textsf{A}(t) = \textsf{A}^\textrm{c}(t) +  \textsf{A}^\textrm{nc} (t)$.
The general solution of Eq.~\eqref{eq:linearEq} is
\begin{align}
	\textbf{\textit{x}}(t) = \textsf{K}^{(t,0)} \textbf{\textit{x}}(0) + \int_0^t ds~ \textsf{K}^{(t,s)} \bm{\zeta}(s) \label{eq:x_solution}
\end{align}
with the propagator $\textsf{K}^{(b,a)} = \exp\left[ -\int_a^b dt~ \textsf{A}(t) \right]$.
As the active noise $\bm{\zeta}$ is non-Markovian in general, we should be careful in preparing the initial condition.
Here we consider the situation that
the system and the environment are separated for $t<0$, and then connected for $t \geq 0$. Therefore, there exists no correlation between the initial position $\textbf{\textit{x}}(0)$ and the initial noise $\bm{\zeta}(0)$, i.e.,  $ \langle \textbf{\textit{x}}^\textsf{T}(0) \bm{\zeta}(0) \rangle = 0$, and then obviously $ \langle \textbf{\textit{x}}^\textsf{T}(0) \bm{\zeta}(t) \rangle = 0$ for $t>0$ with the assumption that  $\bm{\zeta}$ is independent of  position.

The covariance (correlation) matrix of $x_i(t)$ and $x_j(t^\prime)$ can be calculated, using Eq.~\eqref{eq:x_solution}, as
\begin{align}
	\langle x_i(t) x_j(t^\prime) \rangle &= \sum_{kl} \textsf{K}_{ik}^{(t,0)} \textsf{K}_{jl}^{(t^\prime,0)} \langle x_k(0) x_l (0) \rangle \nonumber \\
%	&+ \sum_{kl} \int_0^{t^\prime} ds^\prime K_{ik}^{(t,0)} K_{jl}^{(t^\prime,s^\prime)} \langle x_k (0) \eta_l (s^\prime) \rangle \nonumber \\
%	&+ \sum_{kl} \int_0^{t} ds K_{ik}^{(t,s)} K_{jl}^{(t^\prime,0)}  \langle x_l (0) \eta_k (s) \rangle \nonumber \\
	&+ \sum_{kl} \int_0^{t} ds \int_0^{t^\prime} ds^\prime
	\textsf{K}_{ik}^{(t,s)} \textsf{K}_{jl}^{(t^\prime,s^\prime)}  \langle \zeta_k (s) \zeta_l (s^\prime) \rangle \label{eq:covariance_xx}
\end{align}
and the covariance matrix of $x_i(t)$ and $\eta_j(t^\prime)$ becomes
\begin{align}
	\langle x_i(t) \zeta_j(t^\prime) \rangle = \sum_{k} \int_0^{t} ds~ \textsf{K}_{ik}^{(t,s)}   \langle \zeta_k (s) \zeta_j (t^\prime) \rangle .  \label{eq:covariance_x_eta}
\end{align}
Equations~\eqref{eq:covariance_xx} and \eqref{eq:covariance_x_eta} show that the covariance function depends only on the two-point correlation function of noise, i.e., $\langle \zeta_i (t) \zeta_j (t^\prime)\rangle$, but not on the higher-order multi-point correlations. This means that the covariance function will remain the same even for different active-noise models, as long as their two-point correlations are the same.

We now calculate the average work and heat generated by a linear force in the overdamped dynamics.
First, from Eq.~\eqref{eq:heat_work_def3},the work done by the nonconservative force can be written as
\begin{align}
	\langle W^\textrm{nc} (t) \rangle &=  -\int_0^t ds~ \left\langle \dot{\textit{\textbf{x}}}^\textsf{T} (s)  \circ \Gamma \textsf{A}^\textrm{nc} (s) \textit{\textbf{x}}(s) \right\rangle  \nonumber \\
	& = -\int_0^t ds~ \left\langle \left(-\textsf{A} (s) \textit{\textbf{x}}(s)+ {\bm \zeta}(s) \right)^\textsf{T} \circ \Gamma \textsf{A}^\textrm{nc} (s) \textit{\textbf{x}} (s) \right\rangle \nonumber \\
	& = \int_0^t ds~ \left( \sum_{i,j,k} \textsf{A}_{ij}^\textsf{T} (s) \gamma_j \textsf{A}_{jk}^\textrm{nc} (s) \langle x_i (s) x_k (s) \rangle \right. \nonumber \\
	& ~~~~~~~~~~~~~~~\left.  - \sum_{ij} \gamma_i \textsf{A}_{ij}^\textrm{nc} (s)  \langle \zeta_i (s) \circ x_j (s)\rangle \right). \label{eq:work_nc}
\end{align}
Note that the Stratonovich product between state variables like $x_i$  can be replaced by the Ito product (no extra symbol), but can not be
ignored when the $\delta$- correlated noise $\zeta_i$ is involved. For example, in case of the shot noise in Eq.~\eqref{eq:Shot_corr},
we can easily find that $\langle \zeta_i (s) \circ x_j (s)\rangle =\langle \zeta_i (s)  x_j (s)\rangle +
\delta_{ij}	\lambda_i \langle c^2 \rangle_{p_i}/2$. For the non-Markovian noise with a finite persistence time $\tau$,
the Stratonovich product is the same as the Ito product.

Second, for the Jarzynski work, we define the potential as $U(\textit{\textbf{x}}, \lambda) = \textit{\textbf{x}}^\textsf{T} \textsf{U} \textit{\textbf{x}} $ satisfying  $\Gamma \textit{\textbf{A}}^\textrm{c}(t) \textit{\textbf{x}} = \nabla U(\textit{\textbf{x}}, \lambda)$. Then, from Eq.~\eqref{eq:heat_work_def2}, the Jarzynski work is given by
\begin{align}
	\langle W^\textrm{p} (t) \rangle &= \int_0^t ds \dot{\lambda} \langle \partial_\lambda U  \rangle  = \int_0^t ds ~\dot{\lambda} \sum_{ij} \partial_\lambda \textsf{U}_{ij}(\lambda) \langle x_i (s) x_j(s) \rangle \label{eq:work_c}
\end{align}
Finally, from Eq.~\eqref{eq:heat_work_def1}, the  heat is expressed as
\begin{align}
	\langle Q_i(t) \rangle  &= \int_0^t ds~ \left\langle \dot{x}_i (s)   \circ\gamma_i \textsf{A}_{ij}(s) x_j (s) \right\rangle \nonumber \\
	& = \int_0^t ds \left( -\sum_{j,k} \gamma_i \textsf{A}_{ik}(s)  \textsf{A}_{ij}(s) \langle x_j (s) x_k (s) \rangle \right. \nonumber \\
	&~~~~~~~~~~~~~~\left. + \sum_{j}\gamma_i \textsf{A}_{ij} (s) \langle \zeta_i (s)\circ x_j (s) \rangle \right). \label{eq:heat}
\end{align}
As shown in Eqs.~\eqref{eq:work_nc}, \eqref{eq:work_c}, and \eqref{eq:heat}, the average work and  heat are fully determined by the covariance matrices, and thus by the two-point correlation function $\langle \zeta_i (t) \zeta_j (t^\prime)\rangle$ only.
Therefore, for a linear system, the engine performance is not affected by the non-Gaussianity of a noise which is imbedded in higher-order multi-point (more than two) correlation functions, while it depends on the non-Markovianity characterized by the persistence time $\tau$ defining the two-point correlation function. {We finally note that, though average values of work and heat are independent of model specifics as long as the two-point correlations of noises are the same, higher order quantities such as skewness of the distribution and fluctuation of work and heat are model-dependent~\cite{Roy}.   }

% % % % % % % % % % % % % % % % % %
\begin{figure*}
	\centering
	\includegraphics[width=0.9\linewidth]{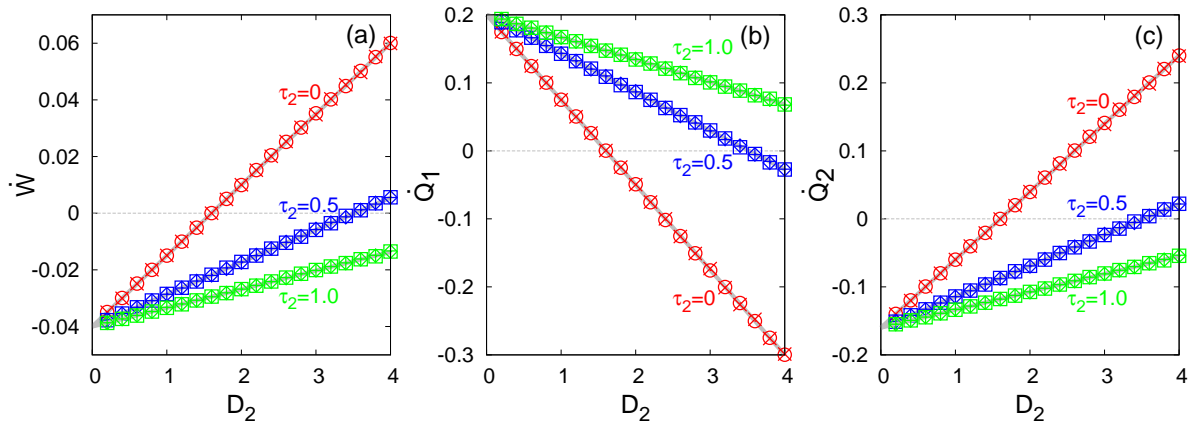}
	\caption{Steady-state rate of the work (a) and  heat (b, c) produced from the system trapped in a harmonic potential and driven by a rotational force. $\bigcirc$ and $\times$ symbols denote data for the equilibrium and the shot noises ($\tau_2=0$), respectively. $+$, $\square$, and $\Diamond$ symbols denote data for the colored-Poisson, the AOUP, and the ABP model, respectively. Gray solid lines represent the analytic results from Eqs.~\eqref{eq:analytic_steady} and \eqref{eq:analytic_steady2}, which are perfectly matched with numerical data. The same values of $\tau_2$ and $D_2$ lead to the same rate regardless of noise models. Parameters used for these simulations are as follows: $\lambda_2$ is varied from $0.2$ to $4$ with $\langle c^2\rangle_{p_2} =2 $ for the shot and the colored-Poisson noise model, $D_2$ is varied from $0.2$ to $4$ for the AOUP model, and 	$\alpha$ is varied from $0.2$ to $4$ with the relation $\alpha=v_0^2\tau/2$ for the ABP model.
	 } \label{fig:fixed_bath_steady}
\end{figure*}
% % % % % % % % % % % % % % % % % % % % %

\subsection{constant reservoirs}
\label{sec:linearB}

In this section, we study an engine driven by a linear force in contact with the (active) reservoirs, of which the noises have the same noise-autocorrelation form in Eq.~\eqref{eq:corr} with two constant parameters; noise strength $D_i$ and persistence time $\tau_i$.
Various noises in Sec.~\ref{sec:various_models} are considered with the same values of $D_i$ and $\tau_i$ with the same initial state
of the engine. Thus, we expect the same values for work and heat, independent of models from Eqs.~\eqref{eq:work_nc},  \eqref{eq:work_c}, and \eqref{eq:heat}.

First, we consider a two-dimensional Brownian particle trapped in a harmonic potential and driven by a nonconservative rotational force. The dynamics is described by Eq.~\eqref{eq:linearEq} with $\Gamma$ and time-independent $\textsf{A}^\textrm{c}$ and $\textsf{A}^\textrm{nc}$ given by
\begin{align}
	\Gamma = \left(
\begin{array}{cc}
	\gamma_1 & 0 \\
	0 & \gamma_2\\
\end{array}
\right),~~~
	\textsf{A}^\textrm{c} = \Gamma^{-1}\left(
  \begin{array}{cc}
  	k & 0 \\
    0 & k\\
  \end{array}
\right),~~~
	\textsf{A}^\textrm{nc} = \Gamma^{-1} \left(
\begin{array}{cc}
	0 & -\epsilon \\
	-\delta& 0\\
\end{array}
\right).
\end{align}
The position coordinate `$x_1$' and `$x_2$' are connected to reservoirs $1$ and $2$ with noises $\zeta_1$ and $\zeta_2$, respectively. In this example, we take reservoir $1$ as an equilibrium reservoir with temperature $T_1$, thus $\langle \zeta_1 (t) \zeta_1 (t^\prime) \rangle = 2 T_1  \gamma_1^{-1} \delta(t-t^\prime)$. Reservoir $2$ is an active reservoir of which the noise satisfies $\langle \zeta_2 (t) \zeta_2 (t^\prime) \rangle = D_2 \tau_2^{-1} \gamma_2^{-2} \exp(-|t-t^\prime|/\tau_2) $. Note that reservoir $2$ can also be an equilibrium reservoir in the $\tau_2 \rightarrow 0$ limit with temperature $T_2 = D_2/\gamma_2$. In this setup, we can obtain the analytic expressions for the average rates of the work and heat in the steady state from Eqs.~\eqref{eq:work_nc} and \eqref{eq:heat}. The results are~\cite{JSLee2020active}
\begin{align}
	\dot{W} = (\delta - \epsilon)\langle x_1 \dot{x}_2 \rangle^\textrm{ss}, ~~\dot{Q}_1 = \epsilon \langle x_1 \dot{x}_2 \rangle^\textrm{ss},~~\dot{Q}_2 = -\delta \langle x_1 \dot{x}_2 \rangle^\textrm{ss}, \label{eq:analytic_steady}
\end{align}
where $\langle \cdots \rangle^\textrm{ss}$ denotes the steady-state average and  $\langle x_1 \dot{x}_2 \rangle^\textrm{ss}$ is given by
\begin{align}
	\langle x_1 \dot{x}_2 \rangle^\textrm{ss} = \frac{1}{k(\gamma_1 +\gamma_2)} \left( \delta T_1 - \frac{\epsilon D_2  }{G_2 \mathcal{B}} \right), \label{eq:analytic_steady2}
\end{align}
with $\mathcal{B} = 1 + k\gamma_2 \tau_2 /(\gamma_1 G_2) + (k^2 - \delta \epsilon) \tau_2^2 /(\gamma_1 G_2)$ and  $G_2= \gamma_2 + k \tau_2 $.

Figure~\ref{fig:fixed_bath_steady} shows the averge rates of work and  heat in the steady state for various models as a function of $D_2$
for $\tau_2=0$, $0.5$, and $1.0$. For these simulations, we set $k=\gamma_1=\gamma_2 = 1$, $T_1 =2$, $\epsilon = 0.5$, and $\delta = 0. 4$. Other parameters of respective models are specified in the caption of Fig.~\ref{fig:fixed_bath_steady}.
Solid curves denote the analytic results in Eqs.~\eqref{eq:analytic_steady} and \eqref{eq:analytic_steady2} and dots are simulation results obtained by averaging over $10^6$ samples in the steady state. The $\tau_2=0$ case corresponds to the $\delta$-correlated noise for reservoir 2, which can be either equilibrium or shot noise reservoir.
The results of these two models coincide with each other and are exactly fitted by the analytic curve for $\dot{W}$, $\dot{Q}_1$, and $\dot{Q}_2$ as shown in Figs. ~\ref{fig:fixed_bath_steady}(a), ~\ref{fig:fixed_bath_steady}(b), and ~\ref{fig:fixed_bath_steady}(c), respectively. The colored-Poisson, the AOUP, and the ABP  models belong to the case with
finite $\tau$.
As seen in Figs. ~\ref{fig:fixed_bath_steady}(a), ~\ref{fig:fixed_bath_steady}(b), and ~\ref{fig:fixed_bath_steady}(c), data points of three different models with the same finite $\tau_2$ coincide with each others and are perfectly matched to a single analytic curve. This clearly demonstrates that the work and  heat are determined solely by the two-point noise correlation function, regardless of other details of the noise statistics.

The second example is a one-dimensional Brownian particle trapped in a breathing harmonic potential. The motion is described by the following equation:
\begin{align}
	\dot{x} = -k(t) x /\gamma + \zeta, \label{eq:breathing}
\end{align}
where $k(t)$  is the time-dependent stiffness with period $\mathcal{T}_\textrm{p}$  given by
\begin{align}
	k(t) = \left\{
	\begin{array}{cc}
		k_\textrm{min} + \omega t, & 0 \leq t < \mathcal{T}_\textrm{p}/2 \\
		k_\textrm{min} + \omega (\mathcal{T}_\textrm{p} - t). & \mathcal{T}_\textrm{p}/2 \leq t < \mathcal{T}_\textrm{p} \\
	\end{array}
	\right.~, \label{eq:k(t)}
\end{align}
where the noise autocorrelation function is given by
$\langle \zeta(t) \zeta(t^\prime) \rangle =   D \tau^{-1} \gamma^{-2} \exp(-|t-t^\prime|/\tau )$. For this simulation, $k_\textrm{min} = \gamma = D = \omega =1 $ and $\mathcal{T}_\textrm{p}=8$ are used and the initial state is set to be $x(0) = 0$. With these conditions, we numerically calculate the accumulated Jarzynski work $W$ from Eq.~\eqref{eq:work_c} and the accumulated heat $Q$ from Eq.~\eqref{eq:heat}.

\begin{figure}
\centering
\includegraphics[width=0.9\linewidth]{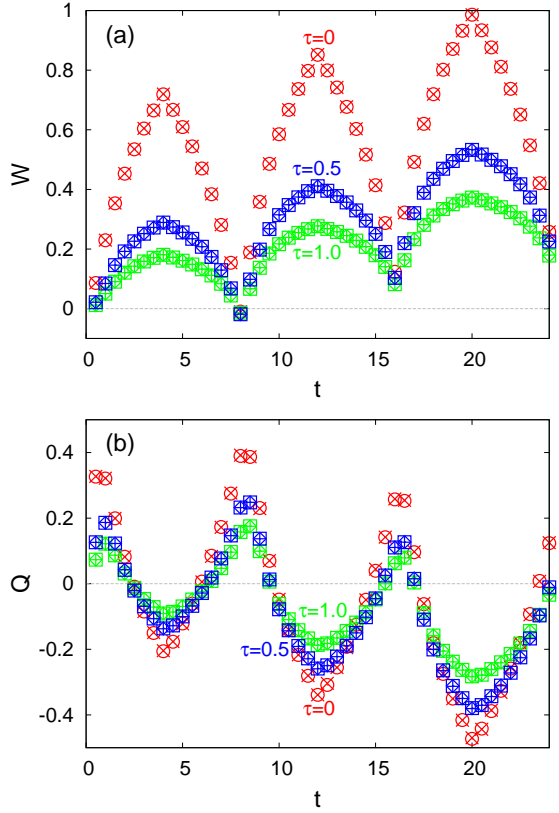}
\caption{Accumulated work (a) and heat (b) produced from the system driven by a periodically changing harmonic force. $\bigcirc$ and $\times$ symbols denote data for the equilibrium and the shot noises ($\tau=0$), respectively. $+$, $\square$, and $\Diamond$ symbols denote data for the colored-Poisson, the AOUP, and the ABP model, respectively. The same values of $\tau$ and $D$ lead to the same value for work and heat regardless of noise models.} \label{fig:fixed_bath_nonsteady}
\end{figure}

Figure~\ref{fig:fixed_bath_nonsteady} shows $W$ and $Q$ as a function of time. Similar to the previous example,
the $\tau=0$ case corresponds to the equilibrium and the shot noise model. Here,  $T=1$ is used for the equilibrium noise and $\lambda = 1$ and $\langle c^2\rangle_{p} =2$ are used for the shot noise. As expected, the two results are exactly matched to each other.
Simulations for the colored-Poisson, the AOUP, and the ABP models are also performed for finite $\tau = 0.5, 1.0$.  In this calculation, $\lambda = 1$, $D =1$, and $\alpha =1$  are used. As the figure shows, the same values of $\tau$ and $D$ lead to the same amount of the work and  heat regardless of noise models. This again demonstrates that the engine performance for a linear system is affected only by the non-Markovianity, but not by the non-Gaussianity,  even for a system driven by an arbitrary time-dependent protocol.

% % % % % % % % % % % % % % % % % %
% % % % % % % % % % % % % % % % % % % % %

\subsection{temporal reservoirs and memory effects}
\label{sec:linearC}

Cyclic protocol of conventional heat engines such as Carnot and Stirling engines is usually accompanied with temporal changes of heat reservoirs. Here, we consider the same temporal changes of active reservoirs. In the recent experiment on the Stirling engine with a bacterial active reservoir~\cite{Krishnamurthy}, the effective temperature of the active reservoir was varied by changing the activity of bacteria by controlling the ambient temperature. As the active particles (bacteria) tend to maintain their motion within a given persistence time, the noise statistics of the active reservoir will not be changed abruptly when external control parameters are changed, but instead have some memory effect originated from its past state. Thus, this reservoir-memory effect should be taken into consideration when an active reservoir changes in time, which has not been recognized and investigated so far in literatures.

% % % % % % % % % % % % % % % % % %
\begin{figure}
	\centering
	\includegraphics[width=0.7\linewidth]{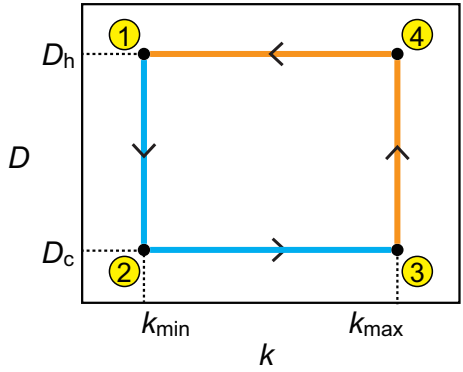}
	\caption{Schematic of the cycle of the Stirling engine. (i) isochoric process $1\rightarrow 2$: $D$ is suddenly switched from $D_\textrm{h}$ to $D_\textrm{c}$ with fixed $k=k_\textrm{min}$ at $t =0$. (ii) isoactive (isothermal) process $2 \rightarrow 3$: $k$ is smoothly changed from $k_\textrm{min}$ to $k_\textrm{max} = k_\textrm{min} + \mathcal{T}_\textrm{p} \omega / 2$ with fixed $D=D_\textrm{c}$ during $0< t <\mathcal{T}_\textrm{p}/2$.  (iii) isochoric process $3 \rightarrow 4$: $D$ is abruptly changed from $D_\textrm{c}$ to $D_\textrm{h}$ with fixed $k=k_\textrm{max}$ at $t =\mathcal{T}_\textrm{p}/2$. (iv) isoactive (isothermal) process $4 \rightarrow 1$: $k$ is smoothly changed from $k_\textrm{max}$ to $k_\textrm{min}$ with fixed $D=D_\textrm{h}$ during $\mathcal{T}_\textrm{p}/2 < t <\mathcal{T}_\textrm{p}$.
 } \label{fig:Stirling_schematic}
\end{figure}
% % % % % % % % % % % % % % % % % % % % %

To investigate the memory effect on the engine performance,  we consider the Stirling engine with an active reservoir. A one-dimensional Brownian particle is trapped in a time-dependent harmonic potential with stiffness $k(t)$ and in contact with a temporal active reservoir with constant persistence time $\tau$ and time-varying noise strength $D(t)$.
For simplicity, we take the time-dependent protocol given by Eqs.~\eqref{eq:breathing} and \eqref{eq:k(t)}.
The cyclic engine protocol consists of four steps which are shown in Fig.~\ref{fig:Stirling_schematic};  (i) isochoric process $1 \rightarrow 2$: $D$ is suddenly switched from $D_\textrm{h}$ to $D_\textrm{c}$ with fixed $k=k_\textrm{min}$ at $t =0$. This process corresponds  to the sudden temperature change of the heat bath with fixed volume of the conventional Stirling engine.
(ii) isoactive process $2 \rightarrow 3$: $k$ changes linearly from $k_\textrm{min}$ to $k_\textrm{max} \equiv k_\textrm{min} + \mathcal{T}_\textrm{p} \omega / 2$ with fixed $D=D_\textrm{c}$ during $0< t <\mathcal{T}_\textrm{p}/2$. This process corresponds to the isothermal compression process of the conventional engine.  (iii) isochoric process $3 \rightarrow 4$: $D$ changes abruptly from $D_\textrm{c}$ to $D_\textrm{h}$ with fixed $k=k_\textrm{max}$ at $t =\mathcal{T}_\textrm{p}/2$. (iv) isoactive process $4 \rightarrow 1$: $k$ changes linearly from $k_\textrm{max}$ to $k_\textrm{min}$ with fixed $D=D_\textrm{h}$ during $\mathcal{T}_\textrm{p}/2 < t <\mathcal{T}_\textrm{p}$. This process corresponds to the isothermal expansion process of the conventional engine.

% % % % % % % % % % % % % % % % % %
\begin{figure*}
	\centering
	\includegraphics[width=0.6\linewidth]{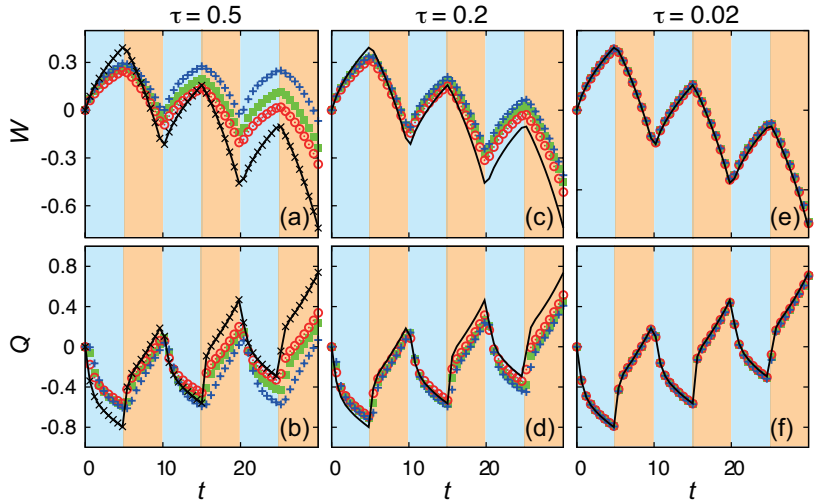}
	\caption{Accumulated work (a, c, e) and heat (b, d, f) produced from the Stirling engine for various $\tau$. $+$, $\blacksquare$, and $\bigcirc$ symbols denote data for the colored-Poisson, the AOUP, and the ABP model with finite persistence time $\tau$, respectively. Solid curves are the numerical results for the equilibrium noise as a reference. $\times$ symbols denote data for the shot noise, which exactly match the equilibrium curve. The colored-Poisson, the AOUP, and the ABP data approach the equilibrium curve as $\tau$ goes to zero.  } \label{fig:Stirling_linear_tau}
\end{figure*}
% % % % % % % % % % % % % % % % % % % % %

% % % % % % % % % % % % % % % % % %
\begin{figure*}
	\centering
	\includegraphics[width=0.6\linewidth]{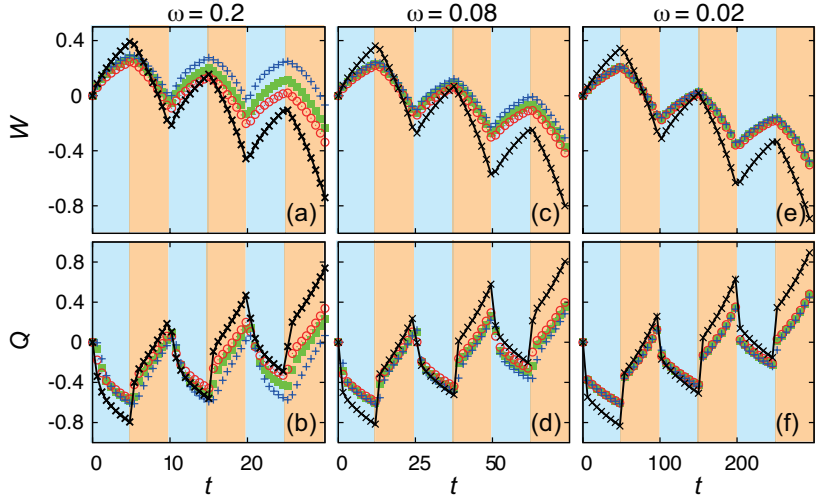}
	\caption{Accumulated work (a, c, e) and heat (b, d, f) produced from the Stirling engine for various protocol speed $\omega$. $+$, $\blacksquare$, and $\bigcirc$ symbols denote data for the colored-Poisson, the AOUP, and the ABP model with finite persistence time $\tau=0.5$, respectively. Solid curves are the numerical results for the equilibrium noise as a reference. $\times$ symbols denote data for the shot noise, which exactly match the equilibrium curve. As $\omega$ goes to zero,
the colored-Poisson, the AOUP, and the ABP data collapse on each other, but not on the equilibrium curve.
} \label{fig:Stirling_linear_omega}
\end{figure*}
% % % % % % % % % % % % % % % % % % % % %

Figure~\ref{fig:Stirling_linear_tau} shows the accumulated work and heat of this Stirling engine for different noise models and different $\tau$. For this simulation, we set the protocol parameters as  $\omega = 0.2 $, $\mathcal{T}_\textrm{p} = 10$, $k_\textrm{min} = 1$ (thus, $k_\textrm{max}=2$) with the reservoir parameters as $\gamma=1$, $D_\textrm{h} = 2$, and $D_\textrm{c} = 1$.
Note that the data for the equilibrium noise and the shot noise (both $\tau=0$) coincide with each other for $W$ and $Q$.
In contrast to these  cases, $W$ and $Q$ data points of the colored-Poisson, the AOUP, and the ABP models with finite $\tau$ do not agree with the others even though they have the same $\tau$ and the same sequence of $D$. This discrepancy
is due to the distinct memory effect for different models, which
becomes smaller as the persistence time $\tau$ decreases as shown in Fig.~\ref{fig:Stirling_linear_tau}. All data eventually
collapse on the equilibrium curve in the $\tau \rightarrow  0$ limit.

% % % % % % % % % % % % % % % % % %
\begin{figure*}
	\centering
	\includegraphics[width=0.9\textwidth]{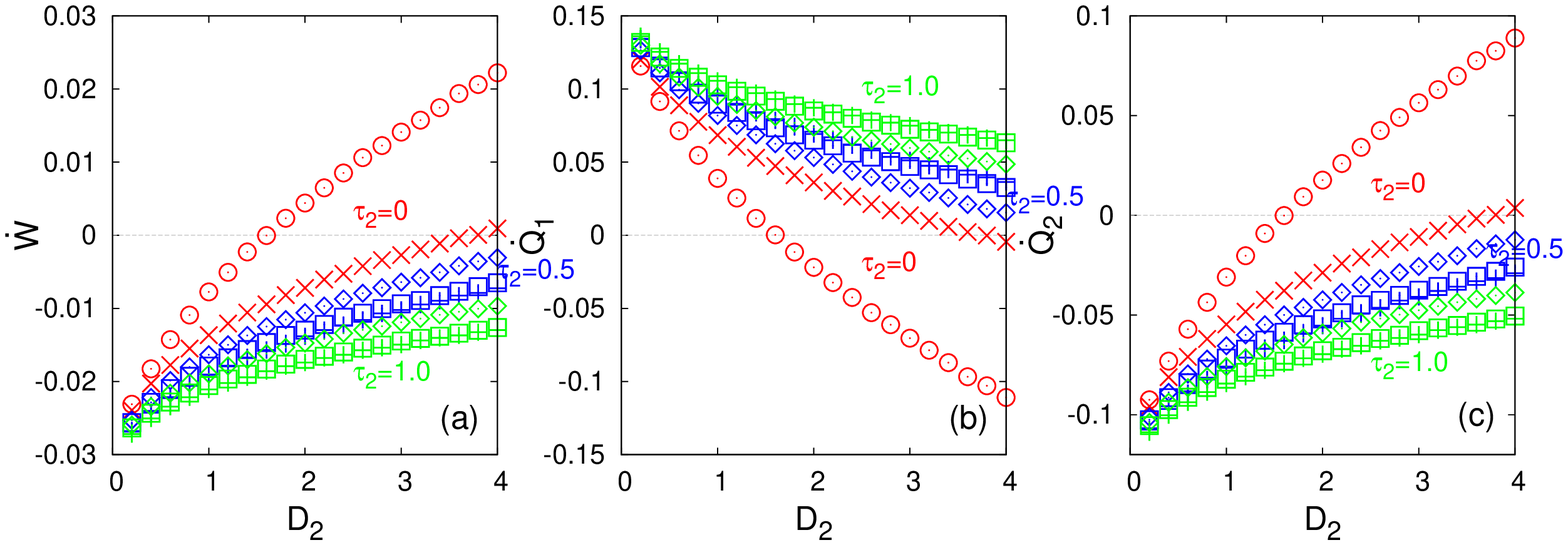}
	\caption{Steady-state rate of the work (a) and heat (b, c) produced from the system trapped in an anharmonic potential with $q=4$ and driven by a rotational force. $\bigcirc$ and $\times$ symbols denote data for the equilibrium and the shot noises ($\tau_2=0$), respectively. $+$, $\square$, and $\Diamond$ symbols denote data for the colored-Poisson, the AOUP, and the ABP model, respectively. Different from the system with a linear force,  the same values of $\tau_2$ and $D_2$ do not yield the same rates. } \label{fig:steady_nonlinear}
\end{figure*}
% % % % % % % % % % % % % % % % % % % % %

In Appendix~\ref{seca:autocorr_var}, we explicitly calculate the noise autocorrelation functions for various active models
when $\tau$ and $D$ is changed abruptly at a certain time $t_\textrm{c}$. In most cases, the noise autocorrelation function involving
a later time than $t_c$ does not maintain the simple exponential form in Eq.~\eqref{eq:corr} and becomes more complicated with the memory effect for finite $\tau$, which depends on the details of the model. This model-dependent memory effect leads to the difference of $W$ and $Q$ for the different active-noise models in Fig.~\ref{fig:Stirling_linear_tau}.
As this difference lasts for several persistence times until a noise is relaxed,  the discrepancy gets smaller  when $\tau$ gets smaller.

One may consider a speed variation of the time-dependent protocol, i.e.~varying $\omega$ and $\mathcal{T}_p$ while keeping the values of $k_\textrm{min}$ and $k_\textrm{max}$. For a very slow process (small $\omega$ or large $\mathcal{T}_p$), the memory effect can
be ignored for $\tau \ll \mathcal{T}_p/2$. We confirm this by numerical simulations.
Figure~\ref{fig:Stirling_linear_omega} shows the plots of $W$ and $Q$ for various protocol speed $\omega = 0.2, 0.08, 0.02$. For relatively high speed ($\omega = 0.2$),  $W$ and $Q$ of the colored-Poisson, the AOUP, and the ABP models are different from the others, while they almost coincide with each other for a very slow process ($\omega = 0.02$). Note that these results can not match the equilibrium data
even in the quasistatic limit ($\omega\rightarrow 0$), due to the finite persistent time $\tau$.

\section{Engine with a nonlinear force}
\label{sec:nonlinear}

\subsection{effect of non-Gaussianity with a nonlinear force} \label{sec:nonlinearA}

When the total mechanical force is nonlinear in position, the work and heat cannot be simply expressed by the two-point noise correlation functions as discussed in Sec.~\ref{sec:linear}, but higher-order correlation functions are necessary in general. Therefore, in this nonlinear case, the non-Gaussianity of a noise should contribute to the work and heat in general.

To investigate the non-Gaussian effect explicitly, we consider a similar steady-state engine studied in Sec.~\ref{sec:linearB}, i.e.~a two-dimensional Brownian particle trapped by an anharmonic potential $U(x_1,x_2)=\frac{k}{q}(x_1^q+x_2^q)$ where $q$ is an even integer,  and driven by a linear nonconservative force ${\textit{\textbf{f}}^\textrm{nc}(\textit{\textbf{x}})}^\textsf{T}= (\epsilon x_2, \delta x_1)$.  Thus, the equation of motion can be written as
\begin{align}
	\gamma_1 \dot{x}_1 &= - k x_1^{q-1} +\epsilon x_2 + \gamma_1\zeta_1~, \nonumber \\
	\gamma_2 \dot{x}_2 &= - k x_2^{q-1} +\delta x_1 + \gamma_2 \zeta_2~,  \label{eq:nonlinear_ss_engine}
\end{align}
where $\zeta_1$ is the equilibrium noise satisfying $\langle \zeta_1 (t) \zeta_1 (t^\prime)\rangle = 2 \gamma_1^{-1} T_1 \delta(t-t^\prime)$ and  $\zeta_2$ is the active noise satisfying $\langle \zeta_2 (t) \zeta_2 (t^\prime)\rangle = D_2 \tau_2^{-1} \gamma_2^{-2} \exp(-|t-t^\prime|/\tau_2)$.

We perform numerical simulations for $q=4$ with the same parameter values  used in Sec.~\ref{sec:linearB}.
Figure~\ref{fig:steady_nonlinear} shows the average rates of the work and heat in the steady state as a function of $D_2$ for various values of $\tau_2$. Notice a clear distinction between data for the equilibrium noise and the shot noise (both $\tau_2=0$), which should
be due to the non-Gaussianity of the shot noise for nonlinear systems. For the other active models with the same $\tau_2$, the simulation data differ from each other as expected.

\subsection{quasistatic process of the Stirling engine with the shot noise}
\label{sec:nonlinearB}

It is interesting to study the quasistatic process of the Stirling engine with the shot noise, subject to the breathing anharmonic potential $U(x)=\frac{k(t)}{q} x^q$ in one dimension.  The equation of motion is given as
\begin{align}
	\dot{x} = -\frac{k(t)}{\gamma} x^{q-1} + \zeta, \label{eq:breathinga}
\end{align}
where the stiffness protocol is given by Eq.~\eqref{eq:k(t)} and $\zeta(t)$ is the shot noise.

Due to the nonlinearity of the mechanical force ($q>2$), we expect that the non-Gaussianity of the shot noise should contribute to the work
and heat as found in the steady-state engine in Sec.~\ref{sec:nonlinearA}.
For example, the Jarzynski work during the process $2\rightarrow 3$ is given from Eq.~\eqref{eq:heat_work_def2} as
\begin{align}
	\langle W_{2 \rightarrow 3} \rangle = \int_{k_\textrm{min}}^{k_\textrm{max}} dk ~ \partial_k \langle U \rangle = \int_{k_\textrm{min}}^{k_\textrm{max}} dk ~ \frac{\langle x^q\rangle}{q} ~.  \label{eq:work2-3.0}
\end{align}
As the system internal energy is given by $\langle U\rangle= \frac{k(t)}{q} \langle x^q \rangle$, the thermodynamic first law guarantees that the heat also depends on $\langle x^q\rangle$.
Hence, the work and heat obviously include the higher-order correlation functions for $q>2$.

However, in the quasistatic limit,
the average $\langle x^q\rangle$ is reduced to a constant independent of the higher-order noise correlation functions, and thus
the non-Gaussianity of the shot noise does not come into play.
In this subsection, we show this interesting result analytically and also perform numerical simulations for $q=4$ and $6$ with various protocol speeds $\omega$.

First, consider the evolution equation of the probability distribution $P(x,t)$, corresponding to Eq.~\eqref{eq:breathinga} as~\cite{entropy}
\begin{align}
	\frac{\partial}{\partial t} P(x,t) &= \frac{\partial}{\partial x} \left( \frac{k(t)}{\gamma} x^{q-1} P(x,t) \right) \nonumber \\
	& + \lambda \int dc p(c) P(x-c,t) - \lambda P(x,t)~, \label{eq:shot_anharmonic_equation}
\end{align}
where $p(c)$ is the distribution function of the shot-noise magnitude $c$ as explained in Sec.~\ref{sec:shot}. Multiplying $x^2$ to Eq.~\eqref{eq:shot_anharmonic_equation} and integrating it over $x$, one can easily find
\begin{align}
	\frac{\partial}{\partial t} \langle x^2 \rangle = -\frac{2 k(t)}{\gamma} \langle x^q \rangle + \lambda \langle c^2\rangle_p. \label{eq:shot_anharmonic_equation2}
\end{align}
In the quasistatic process, the system is almost always in a steady state with the instant value of $k(t)$ at that moment.
In this instant steady state at time $t$, we obtain, by setting both sides of Eq.~\eqref{eq:shot_anharmonic_equation2} zero,
\begin{align}
2 k(t) \langle x^q\rangle^\textrm{ss}=\gamma\lambda\langle c^2\rangle_p~,
\end{align}
which clearly shows that the average $\langle x^q\rangle$ does not depend on the higher-order noise correlations in the quasistatic process.

For later use, we define effective temperatures of the system with the shot-noise reservoir.
From the equipartition relation similar to that in equilibrium, it is reasonable to define for fixed $k$
\begin{align}
	T^\textrm{E} \equiv  k \langle x^q \rangle^\textrm{ss} = \frac{\gamma \lambda}{2} \langle c^2 \rangle_p , \label{eq:shot_steady}
\end{align}
which converges to the conventional temperature $T$ in the equilibrium limit discussed in Sec.~\ref{sec:shot}, i.e.~$\lambda\langle c^2\rangle_p =2 T/\gamma$ fixed in the $\lambda\rightarrow\infty$ limit. One may define another effective temperature
from the diffusive behavior of a particle without any trapping potential ($k=0$) as
\begin{align}
	T^\textrm{D} \equiv \frac{\gamma}{2} \langle x^2 \rangle /t = \frac{\gamma \lambda}{2} \langle c^2 \rangle_p~,
\label{eq:shot_diff}
\end{align}
where Eq.~\eqref{eq:shot_anharmonic_equation2} is used for $k=0$. Notice that
these two effective temperatures are the same for the shot noise, whereas they are different for other active noises in general.

We take the Stirling engine protocol in Fig.~\ref{fig:Stirling_schematic} with the anharmonic potential. For simplicity, we use the temperature notation for the two shot-noise reservoirs with $T_\textrm{c}^\textrm{E}$ and $T_\textrm{h}^\textrm{E}$. Then, in the quasistatic process,
using Eqs.~\eqref{eq:work2-3.0} and \eqref{eq:shot_steady}, we find
\begin{align}
	\langle W_{2 \rightarrow 3} \rangle^\textrm{ss} = \frac{T_\textrm{c}^\textrm{E}}{q} \ln \frac{k_\textrm{max}}{k_\textrm{min}},  \label{eq:work2-3}
\end{align}
which is exactly the same form as the work, Eq.~\eqref{eqa:work1-2}, of the equilibrium-reservoir counterpart explained in Appendix~\ref{seca:Stirling}  by matching
$T_\textrm{c}^\textrm{E}$ with $T_\textrm{c}$. Works and heats for other quasistatic-process segments are also the same as those presented in Eqs.~\eqref{eqa:work_others} and \eqref{eqa:heat}, respectively also by matching
$T_\textrm{h}^\textrm{E}$ with $T_\textrm{h}$.

We define the efficiency of the shot-noise Stirling engine in the conventional way as the ratio of the extracted work versus the heat energy flow from the {\em high-temperature} reservoir:
\begin{align}
	\eta = -\frac{\langle W_{2 \rightarrow 3} \rangle + \langle W_{4 \rightarrow 1} \rangle}{\langle Q_{3 \rightarrow 4} \rangle+ \langle Q_{4 \rightarrow 1} \rangle}.  \label{eq:efficency_Stirling_def}
\end{align}
Then, in the quasistatic process, we obtain
\begin{align}
	\eta_\textrm{E}^\textrm{Stir} = \frac{\eta_\textrm{E}}{1+ \eta_\textrm{E} /\ln \frac{k_\textrm{max}}{k_\textrm{min}} }, ~~\textrm{with }~ \eta_\textrm{E} \equiv 1 - \frac{T_\textrm{c}^\textrm{E}}{T_\textrm{h}^\textrm{E}} \label{eq:efficency_Stirling}
\end{align}
which is the same as that of the equilibrium Stirling engine $\eta_\textrm{C}^\textrm{Stir}$ by replacing $T^\textrm{E}$ with
$T$, see Eq.~\eqref{eqa:efficency_Stirling_equi}. Note that $\eta_\textrm{E}$ is the {effective} Carnot efficiency defined from the effective temperature $T^\textrm{E}$. In the same way, we can define another effective Carnot efficiency $\eta_\textrm{D} \equiv 1- T_\textrm{c}^\textrm{D} / T_\textrm{h}^\textrm{D}$ from $T^\textrm{D}$ defined in Eq.~\eqref{eq:shot_diff}. More discussions on the efficiency of active engines are presented in the next subsection~\ref{sec:nonlinearC}.

% % % % % % % % % % % % % % % % % %
\begin{figure}
	\centering
	\includegraphics[width=0.99\linewidth]{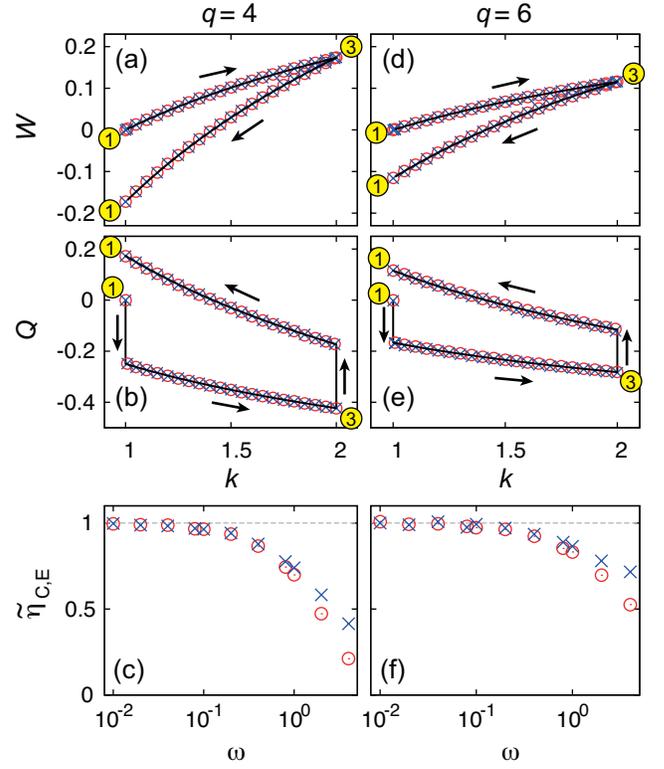}
	\caption{Work, heat, and efficiency of the Stirling engine driven by the time-dependent anharmonic potential. $\bigcirc$ and $\times$ denote data for the equilibrium and the shot noises, respectively. (a) and (b) show the one-cycle accumulated work and heat, respectively, for a very slow (almost quasistatic) process with $\omega = 0.01$ for $q=4$. (c) is the plot of the normalized efficiency  $\tilde{\eta}_\textrm{C}$ and $\tilde{\eta}_\textrm{E}$ as a function of $\omega$ for $q=4$. The efficiency data for the equilibrium and the shot noises coincide with each other for small $\omega$ and show a discrepancy for $\omega\gtrsim 1$.
	(d), (e), and (f) show data for $q=6$.
 } \label{fig:Stirling_anharmonic}
\end{figure}
% % % % % % % % % % % % % % % % % % % % %

Figure~\ref{fig:Stirling_anharmonic} shows the numerical results with various values of the protocol speed $\omega$ for the equilibrium and the shot noises. Accumulated work and heat for $q=4$ are presented in Figs.~\ref{fig:Stirling_anharmonic}(a) and \ref{fig:Stirling_anharmonic}(b), respectively. For these simulations, we set the protocol speed very small as $\omega = 0.01$. Solid curves and data points denote analytic and numerical results, respectively, and they are exactly matched. Figure~\ref{fig:Stirling_anharmonic}(c) shows the plots of the normalized efficiency $\tilde{\eta}$ as a function of $\omega $  with $\tilde{\eta}_\textrm{C} \equiv \eta/ \eta_\textrm{C}^\textrm{Stir}$ for the equilibrium noise and
with $\tilde{\eta}_\textrm{E} \equiv \eta/ \eta_\textrm{E}^\textrm{Stir}$ for the shot noise. As expected, the two efficiencies are almost identical for small $\omega$, but show a discrepancy for $\omega\gtrsim 1$. We find similar results for $q=6$ in Figs.~\eqref{fig:Stirling_anharmonic}(d), \eqref{fig:Stirling_anharmonic}(e) and \eqref{fig:Stirling_anharmonic}(f).
Note that the efficiency of the shot-noise engine is higher than that of the equilibrium engine with a fast protocol speed. This indicates that the efficiency can be enhanced soley by the non-Gaussianity of a noise without any non-Markovianity.

\subsection{steady-state engine}
\label{sec:nonlinearC}

% % % % % % % % % % % % % % % % % %
\begin{figure}
	\centering
	\includegraphics[width=0.99\linewidth]{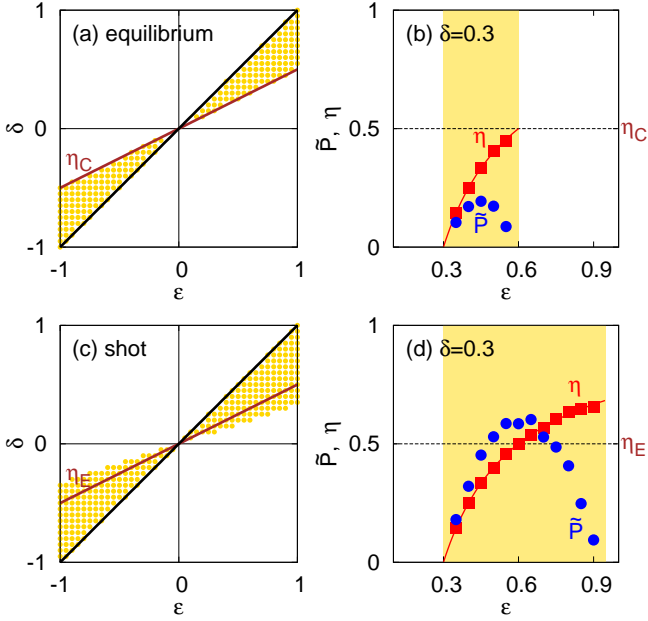}
	\caption{(a) Engine area (yellow-dotted) on the $\epsilon-\delta$ plane of the steady-state engine with an anharmonic potential ($q=4$) and equilibrium reservoirs. The brown line indicates the Carnot-efficiency ($\eta_\textrm{C}$) line. (b) Plot for the efficiency $\eta$ and the normalized power $\tilde{P}$ as a function of $\epsilon$ with fixed $\delta =0.3$. The yellow-shaded area denotes the region satisfying the engine condition. The maximum efficiency is $\eta_\textrm{C}$. (c) Engine area (yellow-dotted) of the steady-state engine ($q=4$) with one equilibrium and one shot-noise reservoir. Note that the engine area is extended over the effective Carnot-efficiency
($\eta_\textrm{E}$) line. (d) Plot for $\eta$ and $\tilde{P} $ as a function of $\epsilon$ with fixed $\delta =0.3$.  The maximum efficiency surpasses $\eta_\textrm{E}$. } \label{fig:efficiency_zero_tau}
\end{figure}
% % % % % % % % % % % % % % % % % % % % %

We investigate the nonlinear effect on the efficiency of the steady-state active engine introduced in
Sec.~\ref{sec:nonlinearA}, where the energy-supplying bath is in equilibrium with temperature $T_1$  and the energy-dissipating bath is an active bath with $D_2$ and $\tau_2$ (effectively low-temperature bath). The efficiency is defined in the conventional way
as the ratio of the total work extraction rate  and the heat flow rate out of the (high-temperature) equilibrium reservoir.

In our previous study~\cite{JSLee2020active}, we investigated the same steady-state active engine model with $q=2$ (linear force) and the AOUP noise. From the study, it was shown that the efficiency can overcome the two effective Carnot efficiencies $\eta_\textrm{D}$ and $\eta_\textrm{E}$ when the persistence times of the two reservoirs are different. Note that surpassing the effective Carnot efficiency in an active engine does not mean the violation of the thermodynamic second law since the dissipation into nonequilibrium reservoirs do not account the full entropy production in general~\cite{JSLee2020active}. For the linear engine, the work and heat are not affected by the non-Gaussianity, thus we expect the same efficiency for other types of active reservoirs as long as the two-point noise correlation functions are identical. That is, surpassing the effective Carnot efficiency can be achieved solely by the non-Markovianity.

However, it is clear that, for a system with a nonlinear force, the non-Gaussianity can also give an additional contribution in enhancing the efficiency. We perform numerical simulations for the steady-state engine described by Eq.~\eqref{eq:nonlinear_ss_engine} for $q=4$
with $T_1 =2$. We use the same values of other parameters as for the linear engine in Sec.~\ref{sec:linearB}, i.e.~$k=\gamma_1=\gamma_2=1$.

Figures~\ref{fig:efficiency_zero_tau}(a) and (b) show the simulation results when the reservoir $2$ is also in equilibrium with temperature $T_2=1$. The yellow dots in Fig.~\ref{fig:efficiency_zero_tau}(a) are plotted on the $\epsilon-\delta$ plane when the model system works as an engine, that is, the engine condition $W<0$ (positive work extraction) and $Q_1>0$ (heat flow out of the reservoir 1) is satisfied.
As expected, the yellow-dotted engine area is restricted in between the two lines: (i) the $\eta=0$ line ($\delta = \epsilon$) and (ii) the $\eta=\eta_\textrm{C}$ line ($\delta = (T_2/T_1) \epsilon$) with the Carnot efficiency $\eta_\textrm{C}=1-T_2/T_1$. Thus, the efficiency is bounded from above by $\eta_\textrm{C}$ as explicitly shown in Fig.~\ref{fig:efficiency_zero_tau}(b), which is the plot for the efficiency and the normalized power $\tilde{P}\equiv {P}/ P_\textrm{eq}^\textrm{max}$ as a function of $\epsilon $ with fixed $\delta = 0.3$.
The {\em global} maximum power is given by $P_\textrm{eq}^\textrm{max}= k T_1 \eta_\textrm{CA}^2 /(\gamma_1 + \gamma_2)$ with $\eta_\textrm{CA} = 1- \sqrt{T_2/T_1}$~\cite{JSLee2020active}.

Figure~\ref{fig:efficiency_zero_tau}(c) shows the engine area when the noise of the reservoir $2$ is the shot noise with $D_2=1$
($\lambda=1$ and  $\langle c^2\rangle_p=2$), leading to $T_2^\textrm{E}=1$. Clearly distinct from Fig.~\ref{fig:efficiency_zero_tau}(a), the engine area is extended over the effective Carnot efficiency line ($\eta=\eta^\textrm{E}$ line). This leads to surpassing the effective Carnot efficiency $\eta_\textrm{E}$ as presented in Fig.~\ref{fig:efficiency_zero_tau}(d). This definitely manifests the non-Gaussian effect on engine efficiency through a nonlinear force without any non-Markovianity.

% % % % % % % % % % % % % % % % % %
\begin{figure}
	\centering
	\includegraphics[width=0.99\linewidth]{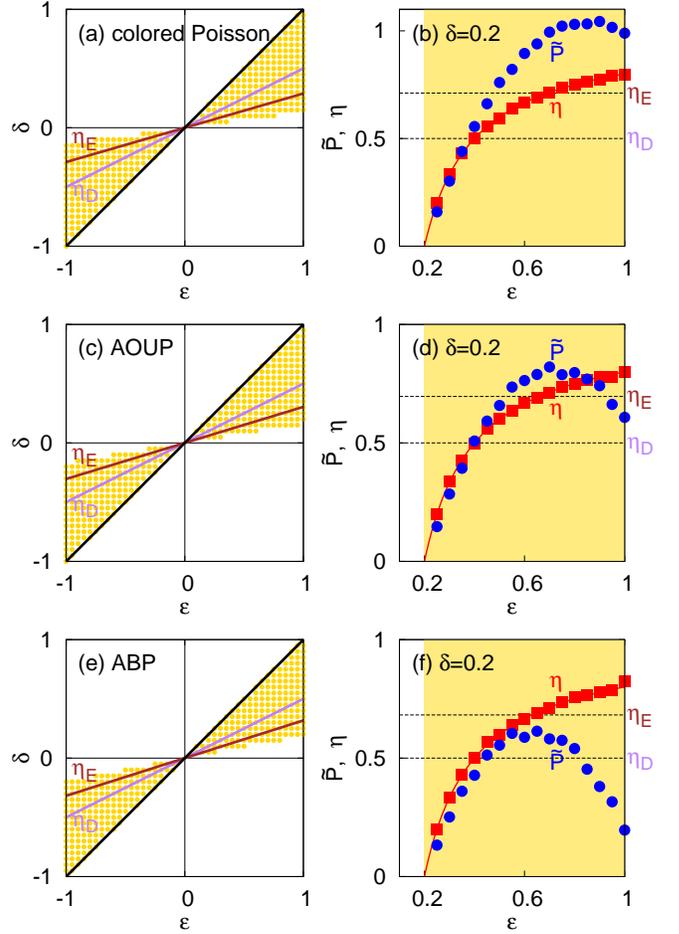}
	\caption{Engine area (yellow-dotted) on the $\epsilon-\delta$ plane of the steady-state engine with an anharmonic potential ($q=4$) with one equilibrium and one (a) colored-Poisson, (c) AOUP, (e) ABP reservoir, respectively.
Note that the engine area is extended over both effective Carnot efficiency ($\eta_\textrm{D}$ and $\eta_\textrm{E}$) lines. (b), (d), and (f) are the corresponding plots for the efficiency $\eta$ and the normalized power $\tilde{P}$ as a function of $\epsilon$ with fixed $\delta =0.3$. The Yellow shaded area denotes the region satisfying the engine condition. The maximum efficiency surpasses both $\eta_\textrm{D}$ and $\eta_\textrm{E}$. } \label{fig:efficiency_nonzero_tau}
\end{figure}
% % % % % % % % % % % % % % % % % % % % %

We also perform similar simulations when the reservoir $2$ generates the colored-Poisson, the AOUP, or the ABP noise with $D_2 = 1$ and $\tau_2 =0.5$. The results are presented in Fig.~\ref{fig:efficiency_nonzero_tau}, which show that the efficiency of all active engines with finite $\tau_2$ can overcome both $\eta_\textrm{D}$ and $\eta_\textrm{E}$. Among these three models, only the AOUP noise is Gaussian. Therefore, one can also conclude that the effective Carnot efficiencies can be overcome solely by the non-Markovian noise without any non-Gaussianity.

\section{Conclusions}
\label{sec:conclusion}

We investigated the effects of the nonequilibrium features of active noise on engine performance using various active noise models. We focused on the non-Gaussianity and non-Markovianity of active noise, and found that the effects could be categorized according to the nature of the mechanical (external) force. First, when the force is linear, the average work and heat are determined only by the two-point noise correlation function. Thus, noise non-Gaussianity is irrelevant to engine performance in such a system. However, non-Makovianity plays an important role in enhancing the engine performance with the efficiency surpassing the effective Carnot efficiency. Furthermore, for a cyclic engine, performance is also affected by the non-Markovian memory of a temporally changing engine environment. Second, when the force is nonlinear, the average work and heat generally depend on higher-order (more than two-point) noise correlation functions, such that non-Gaussianity obviously contributes to engine performance. Thus, non-Gaussianity can enhance engine performance in a nonlinear system. This effect was not well-documented in previous studies; it is relatively difficult to analyze a nonlinear system. We expect that more interesting results might be obtained by studying a general nonlinear system with an active reservoir.

\begin{acknowledgments}
	Authors acknowledge the Korea Institute for Advanced Study for providing computing resources (KIAS Cen- ter for Advanced Computation Linux Cluster System). This research was supported by the NRF Grant No. 2017R1D1A1B06035497 (HP) and the KIAS individual Grants No. PG013604 (HP), PG064901 (JSL) at Korea Institute for Advanced Study.
\end{acknowledgments}

\appendix

\section{Autocorrelation function}

\subsection{colored-Poisson noise}
\label{seca:autocorr_CPoi}
We consider a one-dimensional colored-Poisson noise for simplicity, and thus the index $i$ in Eq.~\eqref{eq:CPoisson_noise} is dropped. The autocorrelation function of the noise $\langle \zeta(t_a)\zeta(t_b) \rangle$ for $t_b\geq t_a$ can be written as
\begin{align}
	\langle \zeta (t_a) \zeta (t_b) \rangle & =  \frac{1}{\tau^2}\left\langle \sum_{n,m} c_{n} c_{m} H(t_a-t_{n}) H(t_b-t_{m}) e^{-\frac{t_a - t_n}{\tau}} e^{-\frac{t_b - t_m}{\tau}} \right\rangle \nonumber \\
	& = \frac{\langle c^2\rangle_p}{\tau^2}\left\langle \sum_{n}   H(t_a-t_{n})  e^{-\frac{t_a + t_b}{\tau}+\frac{2 t_n }{\tau}} \right\rangle~, %\nonumber \\
%	& + \frac{1}{\tau^2}\left\langle \sum_{n \neq m} c_{n} c_{m} H(t_a-t_{n}) H(t_b-t_{m}) e^{-\frac{t_a - t_n}{\tau}} e^{-\frac{t_b - t_m}{\tau}} \right\rangle .
	 \label{eqa:CPoisson_noise}
\end{align}
where we used the noise magnitude correlations as
$\langle c_n c_m\rangle_p = \langle c^2\rangle_p ~\delta_{nm}$ for a given time sequence of shots.
The average over time sequences of shots yields
\begin{align}
	\langle \zeta (t_a) \zeta (t_b) \rangle
	& = \frac{\langle c^2 \rangle_p}{\tau^2}e^{-\frac{t_a + t_b}{\tau}}\left\langle \sum_{n}   H(t_a-t_{n})  e^{\frac{2 t_n }{\tau}} \right\rangle \nonumber \\
	& = \frac{\langle c^2 \rangle_p}{\tau^2}e^{-\frac{t_a + t_b}{\tau}} \int_0^{t_a} e^{\frac{2 t}{\tau} }\lambda dt \nonumber \\
	& = \frac{\lambda \langle c^2 \rangle_p}{2\tau} \left( e^{-\frac{t_b - t_a}{\tau}}- e^{-\frac{t_b+t_a}{\tau}} \right).
	\label{eqa:CPoisson_noise1}
\end{align}
The second equality in Eq.~\eqref{eqa:CPoisson_noise1} comes from the fact that the probability observing a Poisson shot during $dt$ is simply $\lambda dt$. For large $t_a/\tau$, the noise autocorrelation function becomes the same as Eq.~\eqref{eq:Colored_corr}.

\subsection{ABP noise}
\label{seca:autocorr_ABP}
For the ABP model, the self-propulsion force acts as an active noise, i.e., ${\bm \zeta} =  v_0 \textbf{\textit{e}}_\theta$.
In order to derive the noise-autocorrelation function $\langle \zeta_i (t_a) \zeta_j (t_b)\rangle_\theta $, it is necessary to calculate $\langle \cos \theta_{t_a} \cos \theta_{t_b}\rangle_\theta$, $\langle \cos \theta_{t_a} \sin\theta_{t_b}\rangle_\theta$, and $\langle \sin \theta_{t_a} \sin \theta_{t_b}\rangle_\theta$, where $\langle \cdots \rangle_\theta$ denotes the average over $\theta$.  First, $\langle \cos \theta_{t_a} \cos \theta_{t_b}\rangle_\theta$ for $t_b >t_a$ is
\begin{widetext}
\begin{align}
	 \langle \cos \theta_{t_a} \cos \theta_{t_b} \rangle_\theta
	& = \int_0^{2\pi} d \theta_0 \int_{-\infty}^\infty d\theta_{t_a}  \int_{-\infty}^\infty d\theta_{t_b}  \cos \theta_{t_b} P_{t_b - t_a}(\theta_{t_b}| \theta_{t_a})  \cos \theta_{t_a} P_{t_a} (\theta_{t_a} | \theta_0) P^\textrm{init} (\theta_0), \label{eqa:cos_corr}
\end{align}
where $\theta_0$ is the initial value of $\theta$ at time $t=0$, $P_{t_2 -t_1} (\theta_{t_2}|\theta_{t_1})$ is the conditional transition probability observing the processes of which the final state is $\theta_{t_2}$ at time $t_2$ and the initial state is $\theta_{t_1}$ at time $t_1$, and $P^\textrm{init}(\theta_0)$ is the initial distribution of $\theta_0$. From the equation of  motion of $\theta$ in Eq.~\eqref{eq:overLangevinABP}, the conditional probability for $t_2 > t_1$ is given by
\begin{align}
	P_{t_2 -t_1} (\theta_{t_2}| \theta_{t_1} ) = \frac{1}{\sqrt{4\pi D_\theta (t_2-t_1)}} \exp\left[-\frac{(\theta_{t_2} -\theta_{t_1})^2}{4 \pi D_\theta (t_2-t_1) }\right]. \label{eqa:transitionProb}
\end{align}
For simplicity, we take the uniform initial distribution, i.e., $P^\textrm{init} (\theta_0) = 1/2\pi$. Then, Eq.~\eqref{eqa:cos_corr} becomes
\begin{align}
	\langle \cos \theta_{t_a} \cos \theta_{t_b} \rangle_\theta
	& = \int_0^{2\pi} d \theta_0 \int_{-\infty}^\infty d\theta_{t_a}  \int_{-\infty}^\infty d\theta_{t_b}
	\frac{1}{2} \textrm{Re} \left[ e^{i(\theta_{t_a} + \theta_{t_b} )} + e^{i(\theta_{t_a} - \theta_{t_b} ) } \right] P_{t_b - t_a}(\theta_{t_b}| \theta_{t_a})   P_{t_a} (\theta_{t_a} | \theta_0) P^\textrm{init} (\theta_0) \nonumber \\
	& = \frac{1}{2} \int_0^{2\pi} d \theta_0 P^\textrm{init} (\theta_0) \left[ e^{-D_\theta |t_b - t_a|} + e^{2i \theta_0} e^{-D_\theta (4 t_a + |t_b - t_a| )} \right]  = \frac{1}{2} e^{-|t_b - t_a|/\tau}. \label{eqa:ABPnoise_corr_cal}
\end{align}
\end{widetext}
In the similar way, we can also show that $\langle \sin \theta_t  \sin \theta_{t^\prime} \rangle  =  e^{-|t^\prime - t|/\tau}/2$ and $ \langle \cos \theta_t  \sin \theta_{t^\prime} \rangle  = 0$. These results lead to Eq.~\eqref{eq:ABPcorr}.

\section{Autocorrelation function with abrupt changes of $\tau$ and $D$}
\label{seca:autocorr_var}

We consider the case where $\tau$ and $D$ of the one-dimensional active noise is changed abruptly at $t=t_c$ as
\begin{align}
	\tau = \tau_1,&~~D=D_1 ~~~~\textrm{for } t \leq t_c \nonumber \\
	\tau = \tau_2, &~~D=D_2 ~~~~\textrm{for } t > t_c.
	\label{eqa:tau_and_D}
\end{align}
Here we calculate the noise-autocorrelation function $\langle \zeta(t_a) \zeta(t_b) \rangle$ of various active noises for two cases: The first is $t_a < t_c < t_b$ and the second is $t_c < t_a < t_b$.

\subsection{colored-Poisson noise}
\label{seca:autocorr_CPoi_bath_var}

For the colored-Poisson noise, the noise strength is given by $D=\gamma^2\lambda\langle c^2\rangle_p /2$.
We choose $\lambda = \lambda_1$ and $p(c) = p^{(1)}(c)$ for $t \leq t_c $, and $\lambda = \lambda_2$ and
$p(c) = p^{(2)}(c)$ for $t > t_c $.

For $t_a < t_c < t_b$, the noise is written as
\begin{align}
	\zeta (t_a)&=  \sum_n \frac{c_{n}}{\tau_1} H(t_a-t_{n}) e^{-\frac{t_a-t_{n}}{\tau_1}}, \nonumber \\
	\zeta (t_b)&=  \sum_n \frac{c_{n}}{\tau_1} H(t_c-t_{n}) e^{-\frac{t_b-t_{n}}{\tau_1}} \nonumber \\
	&+ \sum_n \frac{c_{n}}{\tau_2} H(t_b-t_{n}) H(t_{n}-t_c) e^{-\frac{t_b-t_{n}}{\tau_2}}.
	 \label{eqa:CPoi_noise1}
\end{align}
Then, the noise autocorrelation function becomes
\begin{widetext}
\begin{align}
	\langle \zeta (t_a) \zeta (t_b) \rangle & =  \frac{1}{\tau_1^2}\left\langle \sum_{n,m} c_{n} c_{m} H(t_a-t_{n}) H(t_c-t_{m}) e^{-\frac{t_a - t_n}{\tau_1}} e^{-\frac{t_b - t_m}{\tau_1}} \right\rangle
	+ \frac{1}{\tau_1 \tau_2}\left\langle \sum_{n,m} c_{n} c_{m} H(t_a-t_{n}) H(t_b-t_{m}) H(t_m-t_{c}) e^{-\frac{t_a - t_n}{\tau_1}} e^{-\frac{t_b - t_m}{\tau_2}} \right\rangle  \nonumber\\ %\label{eqa:CPoi_corr_time_dep1}
& =  \frac{1}{\tau_1^2}\left\langle \sum_{n} c_{n}^2  H(t_a-t_{n})  e^{-\frac{t_a + t_b}{\tau_1}+\frac{2 t_n }{\tau_1}} \right\rangle
	= \frac{\lambda_1 \langle c^2 \rangle_{p^{(1)}} }{2\tau_1} \left( e^{-\frac{t_b - t_a}{\tau_1}} - e^{-\frac{ t_b+ t_a}{\tau_1}} \right),
	\label{eqa:CPoi_corr_time_dep2}
\end{align}
where we used $\langle c_n c_m\rangle_p = \langle c^2\rangle_p ~\delta_{nm}$.
For large $t_a/\tau_1$, the noise autocorrelation function becomes the simple exponential form as in Eq.~\eqref{eq:Colored_corr}.

However, for $t_c < t_a < t_b$, the autocorrelation function has a different form. In this case, the noise is written as
\begin{align}
	\zeta (t_a)&=  \sum_n \frac{c_{n}}{\tau_1} H(t_c-t_{n}) e^{-\frac{t_a-t_{n}}{\tau_1}}+ \sum_n \frac{c_{n}}{\tau_2} H(t_a-t_{n}) H(t_{n}-t_c) e^{-\frac{t_a-t_{n}}{\tau_2}} \nonumber \\	
	\zeta (t_b)&=  \sum_n \frac{c_{n}}{\tau_1} H(t_c-t_{n}) e^{-\frac{t_b-t_{n}}{\tau_1}} + \sum_n \frac{c_{n}}{\tau_2} H(t_b-t_{n}) H(t_{n}-t_c) e^{-\frac{t_b-t_{n}}{\tau_2}}.
	\label{eqa:CPoi_noise2}
\end{align}
Then, we have
\begin{align}
		\langle \zeta (t_a) \zeta (t_b) \rangle & =  \frac{1}{\tau_1^2}\left\langle \sum_{n} c_{n}^2 H(t_c-t_{n})  e^{-\frac{t_a +t_b}{\tau_1} + \frac{ 2t_n}{\tau_1}} \right\rangle
	+ \frac{1}{\tau_2^2}\left\langle \sum_{n} c_{n}^2  H(t_a-t_{n}) H(t_b-t_{n}) H(t_n-t_{c}) e^{-\frac{t_a +t_b }{\tau_2}+ \frac{2 t_n}{\tau_2}}  \right\rangle \nonumber \\
	& = \frac{\lambda_1 \langle c^2 \rangle_{p^{(1)}}}{2\tau_1} \left( e^{-\frac{t_b + t_a-2t_c}{\tau_1}}  - e^{-\frac{ t_b+t_a}{\tau_1}} \right)
	+ \frac{\lambda_2 \langle c^2 \rangle_{p^{(2)}}}{2\tau_2} \left( e^{-\frac{t_b - t_a}{\tau_2}} - e^{-\frac{ t_b+t_a-2t_c}{\tau_2}} \right)
	\label{eqa:CPoi_corr_time_dep3}
\end{align}
Even for large $t_a/\tau_1$, the correlation function does not return to the original simple exponential form.

\subsection{AOUP noise}
\label{seca:autocorr_AOUP_bath_var}

We first consider the case of $t_a < t_c <t_b$.
From Eq.~\eqref{eq:AOUP}, the noises at $t=t_a$ and $t_b$ are given as
\begin{align}
	\zeta(t_a) &=  e^{-\frac{t_a}{\tau_1}} \zeta(0) +  \int_0^{t_a} dt^\prime e^{-\frac{t_a - t^\prime}{\tau_1}} \frac{\sqrt{2 D_1}}{\tau_1 \gamma} \xi(t^\prime), \label{eqa:AOUP_noise_ta}  \\
	\zeta(t_b) &=  e^{-\frac{t_b-t_c}{\tau_2}} \zeta(t_c) +  \int_{t_c}^{t_b} dt^\prime e^{-\frac{t_b - t^\prime}{\tau_2}} \frac{\sqrt{2 D_2}}{\tau_2 \gamma} \xi(t^\prime). \label{eqa:AOUP_noise_tb}
\end{align}
Note that $\zeta(t_c)$ is obtained by substituting $t_a$ with $t_c$ in Eq.~\eqref{eqa:AOUP_noise_ta}.
Multiplying Eqs.~\eqref{eqa:AOUP_noise_ta} and \eqref{eqa:AOUP_noise_tb}, we find
\begin{align}
	\langle \zeta(t_a) \zeta(t_b)  \rangle &= e^{-\frac{t_a + t_c}{\tau_1}} e^{-\frac{t_b - t_c}{\tau_2}} \langle \zeta(0)^2 \rangle \nonumber \\
	&+ e^{-\frac{t_b - t_c}{\tau_2} } \int_0^{t_a} dt^\prime \int_0^{t_c} d t^{\prime\prime} e^{-\frac{t_a - t^\prime}{\tau_1}} e^{-\frac{t_c - t^{\prime \prime} }{\tau_1}} \frac{2D_1}{\tau_1^2 \gamma^2} \langle \xi(t^\prime) \xi(t^{\prime \prime}) \rangle
	+  \int_0^{t_a} dt^\prime \int_{t_c}^{t_b} d t^{\prime\prime} e^{-\frac{t_a - t^\prime}{\tau_1} } e^{-\frac{t_b - t^{\prime \prime} }{\tau_2}}  \frac{2\sqrt{D_1 D_2 } }{\tau_1 \tau_2 \gamma^2} \langle \xi(t^\prime) \xi(t^{\prime \prime}) \rangle. \label{eqa:AOUP_corr1}
\end{align}
By using $\langle \xi(t) \xi(t^\prime)\rangle = \delta(t-t^\prime) $, Eq.~\eqref{eqa:AOUP_corr1} becomes
\begin{align}
	\langle \zeta(t_a) \zeta(t_b)  \rangle = \frac{D_1}{\tau_1 \gamma^2} e^{-\frac{t_c - t_a}{\tau_1} - \frac{t_b - t_c}{\tau_2} } + \left( \langle \zeta(0)^2 \rangle  -\frac{D_1}{\tau_1\gamma^2} \right) e^{-\frac{t_c + t_a}{\tau_1} -\frac{t_b - t_c}{\tau_2}}.  \label{eqa:AOUP_corr2}
\end{align}
For large $t_a /\tau_1$, only the first term of the right hand side in Eq.~\eqref{eqa:AOUP_corr1} survives, which is different from the simple exponential form in Eq.~\eqref{eq:corr}.

For $t_c < t_a < t_b$, $\zeta(t_a)$ is written as
\begin{align}
	\zeta(t_a) &=  e^{-\frac{t_a - t_c}{\tau_2}} \zeta(t_c) +  \int_{t_c}^{t_a} dt^\prime e^{-\frac{t_a - t^\prime}{\tau_2}} \frac{\sqrt{2 D_2}}{\tau_2 \gamma} \xi(t^\prime)~, \label{eqa:AOUP_noise_ta1}
\end{align}
with $\zeta(t_b)$ in Eq.~\eqref{eqa:AOUP_noise_tb}.
Then, the autocorrelation function becomes
\begin{align}
	&\langle \zeta(t_a) \zeta(t_b)  \rangle = \frac{D_1}{\tau_1 \gamma^2} e^{-\frac{ t_b+t_a-2t_c}{\tau_2}}
	+\frac{D_2}{\tau_2 \gamma^2} e^{-\frac{ t_b-t_a}{\tau_2}}
	+ \left( \langle \zeta(0)^2 \rangle  -\frac{D_1}{\tau_1\gamma^2} \right) e^{-\frac{2 t_c}{\tau_1}-\frac{t_a + t_b - 2t_c}{\tau_2}} -\frac{D_2}{\tau_2\gamma^2} e^{-\frac{ t_b+t_a}{\tau_2}}.
	\label{eqa:AOUP_corr3}
\end{align}
For large $t_c /\tau_1$, the first two terms survive.

\subsection{ABP noise}
\label{seca:autocorr_ABP_bath_var}

The noise strength is given by $D=\gamma^2 v_0^2 \tau /2$.
We choose $v_0=v_1$ and $\tau=\tau_1$ for $t\leq t_c$, and $v_0 = v_2$ and $\tau=\tau_2$ for $t > t_c$.
For $t_a < t_c < t_b$, the noise autocorrelation can be written as
\begin{align}
	 \langle v_1 \cos \theta_{t_a} v_2\cos \theta_{t_b} \rangle_\theta
	& = v_1 v_2\int_0^{2\pi} d \theta_0 \int_{-\infty}^\infty d\theta_{t_a} \int_{-\infty}^\infty d\theta_{t_c} \int_{-\infty}^\infty d\theta_{t_b}  \cos \theta_{t_b} P_{t_b - t_c}(\theta_{t_b}| \theta_{t_c}) P_{t_c - t_a}(\theta_{t_c}| \theta_{t_a})  \cos \theta_{t_a} P_{t_a} (\theta_{t_a} | \theta_0) P^\textrm{init} (\theta_0). \label{eqa:ABP_corr_cal2}
\end{align}
Using  \eqref{eqa:transitionProb}, we can easily get
\begin{align}
	\langle v_1 \cos \theta_{t_a} v_2\cos \theta_{t_b} \rangle_\theta = \frac{v_1 v_2}{2} e^{-\frac{t_b - t_c}{\tau_2}} e^{-\frac{t_c - t_a}{\tau_1}}
\end{align}
which is different from  Eq.~\eqref{eq:ABPcorr}.
For $t_c < t_a < t_b$, we find
	\begin{align}
		\langle v_2 \cos \theta_{t_a} v_2\cos \theta_{t_b} \rangle_\theta
		& = v_2^2 \int_0^{2\pi} d \theta_0  \int_{-\infty}^\infty d\theta_{t_c} 		\int_{-\infty}^\infty d\theta_{t_a}		\int_{-\infty}^\infty d\theta_{t_b}  \cos \theta_{t_b} P_{t_b - t_a}(\theta_{t_b}| \theta_{t_a}) \cos \theta_{t_a} P_{t_a - t_c}(\theta_{t_a}| \theta_{t_c})   P_{t_c} (\theta_{t_c} | \theta_0) P^\textrm{init} (\theta_0) \nonumber \\
		& = \frac{v_2^2}{2} e^{-\frac{t_2 -t_1}{\tau_2}}~,
		\label{eqa:ABP_corr1_cal3}
	\end{align}
\end{widetext}
which has the same form as in Eq.~\eqref{eq:ABPcorr}.

\section{Stirling engine using an anharmonic potential with an equilibrium temporal bath} \label{seca:Stirling}

Here, we evaluate the work, the heat, and the efficiency of the one-dimensional Stirling engine using a breathing anharmonic potential $U(x) = k(t) x^q / q$ ($q = 2, 4, 6, \cdots$) with an equilibrium bath. The equation of motion of this engine during the isothermal processes is given by
\begin{align}
	\dot{x} = -\frac{k(t)}{\gamma} x^{q-1} + \zeta, \label{eqa:breathing}
\end{align}
where $k(t)$  is the time-dependent stiffness with period $\mathcal{T}_\textrm{p}$ given by Eq.~\eqref{eq:k(t)} and $\zeta (t)$ is a Gaussian white noise satisfying $\langle \zeta(t) \zeta(t^\prime) \rangle = 2  T(t) \gamma^{-1} \delta(t-t^\prime) $ with $D(t) = \gamma T(t)$. The reservoir protocol is shown in Fig.~\ref{fig:Stirling_schematic}; $T= T_\textrm{c}$ for $0\leq t \leq \mathcal{T}_\textrm{p}/2$  and $T= T_\textrm{h}$ for $\mathcal{T}_\textrm{p}/2 \leq t \leq \mathcal{T}_\textrm{p}$ with the condition $T_\textrm{h} > T_\textrm{c}$. Temperature abruptly changes from $T_\textrm{h}$ to $T_\textrm{c}$ at $t=0$ and from $T_\textrm{c}$ to $T_\textrm{h}$ at $t = \mathcal{T}_\textrm{p}/2$.

For the analytic calculation, we consider a quasistatic process, thus, the engine processes $2\rightarrow 3$ and $4\rightarrow 1$ are very slow ($\omega\rightarrow 0$ limit) and thus are always in equilibrium steady states. Then, the work done by the protocol $k(t)$ from $2$ to $3$ is given by (see Eq.~\eqref{eq:heat_work_def2})
\begin{align}
	\langle W_{2 \rightarrow 3} \rangle^\textrm{ss} = \int_{k_\textrm{min}}^{k_\textrm{max}} dk ~ \partial_k \langle U \rangle^\textrm{ss} = \int_{k_\textrm{min}}^{k_\textrm{max}} dk ~ \frac{\langle x^q\rangle^\textrm{ss}}{q} =\frac{T_\textrm{c}}{q} \ln \frac{k_\textrm{max}}{k_\textrm{min}}~,  \label{eqa:work1-2}
\end{align}
where the equipartition relation as $ k \langle x^q \rangle^\textrm{ss} =T $ in equilibrium at temperature $T$ is used.
Similarly, we obtain
\begin{align}
		\langle W_{4 \rightarrow 1} \rangle^\textrm{ss}  &=  \frac{T_\textrm{h}}{q} \ln \frac{k_\textrm{min}}{k_\textrm{max}}, \nonumber \\
		\langle W_{1 \rightarrow 2} \rangle^\textrm{ss}  &=\langle W_{3 \rightarrow 4} \rangle^\textrm{ss}  = 0.  \label{eqa:work_others}
\end{align}
We evaluate the heat using the thermodynamic first law in Eq.~\eqref{eq:first_law}. As the system internal energy is simply given by
the average of the potential energy $\langle U\rangle^\textrm{ss}=T/q$ in equilibrium, we find the average heat for each process as
\begin{align}
	\langle Q_{1 \rightarrow 2} \rangle^\textrm{ss} &= \langle \Delta U_{1 \rightarrow 2} \rangle^\textrm{ss} - \langle W_{1 \rightarrow 2} \rangle^\textrm{ss} = \frac{1}{q} (T_\textrm{c} - T_\textrm{h}), \nonumber \\
	\langle Q_{2 \rightarrow 3} \rangle^\textrm{ss} &= \langle \Delta U_{2 \rightarrow 3} \rangle^\textrm{ss} - \langle W_{2 \rightarrow 3} \rangle^\textrm{ss} = - \frac{T_\textrm{c}}{q} \ln \frac{k_\textrm{max}}{k_\textrm{min}}, \nonumber \\
	\langle Q_{3 \rightarrow 4}
	\rangle^\textrm{ss} &= \langle \Delta U_{3 \rightarrow 4} \rangle^\textrm{ss} - \langle W_{3 \rightarrow 4} \rangle^\textrm{ss} = \frac{1}{q} (T_\textrm{h} - T_\textrm{c}), \nonumber \\
	\langle Q_{4 \rightarrow 1} \rangle^\textrm{ss} &= \langle \Delta U_{4 \rightarrow 1} \rangle^\textrm{ss} - \langle W_{4 \rightarrow 1} \rangle^\textrm{ss} = - \frac{T_\textrm{h}}{q} \ln \frac{k_\textrm{min}}{k_\textrm{max}}. \label{eqa:heat}
\end{align}
Then, the efficiency of the quasistatically operating Stirling engine using the anharmonic potential with an equilibrium reservoir becomes
\begin{align}
	\eta_\textrm{C}^\textrm{Stir} \equiv -\frac{\langle W_{2 \rightarrow 3} \rangle^\textrm{ss}+ \langle W_{4 \rightarrow 1} \rangle^\textrm{ss}}{\langle Q_{3 \rightarrow 4} \rangle^\textrm{ss}+ \langle Q_{4 \rightarrow 1} \rangle^\textrm{ss}} = \frac{\eta_\textrm{C}}{1+ \eta_\textrm{C} /\ln \frac{k_\textrm{max}}{k_\textrm{min}} }, \label{eqa:efficency_Stirling_equi}
\end{align}
with the conventional Carnot efficiency  $\eta_\textrm{C} =1- T_\textrm{c}/T_\textrm{h} $. Note that this efficiency is independent of $q$.

\vfil\eject

\end{document}